\documentclass[a4paper, journal]{IEEEtran}
\usepackage{url}
\usepackage[utf8]{inputenc}
\usepackage{xcolor}
\usepackage{amsmath}
\usepackage{amssymb}
\usepackage{algorithm}
\usepackage{algorithmic}

\usepackage[acronyms,nonumberlist,nopostdot,nomain,nogroupskip]{glossaries}
\usepackage{tablefootnote}
\usepackage{booktabs}
\usepackage{tabularx}
\usepackage{tikz}
\usepackage{pgfplots}
\pgfplotsset{compat=newest}
\pgfplotsset{plot coordinates/math parser=false}
\newlength\fheight
\newlength\fwidth
\usetikzlibrary{plotmarks,patterns,decorations.pathreplacing,backgrounds,calc,arrows,arrows.meta,spy,matrix,backgrounds}
\usepgfplotslibrary{patchplots,groupplots}
\usepackage{tikzscale}
\usepackage{hyperref}

\usepackage{multirow}
\usepackage{tkz-kiviat}

\usepackage[font=small]{subcaption}
\usepackage[font=small]{caption}

\usepackage{mathtools}

\usepackage{dblfloatfix}    
\usepackage{colortbl}

\newacronym{3gpp}{3GPP}{3rd Generation Partnership Project}
\newacronym{adc}{ADC}{Analog to Digital Converter}
\newacronym{5g}{5G}{5th generation}
\newacronym{aimd}{AIMD}{Additive Increase Multiplicative Decrease}
\newacronym{am}{AM}{Acknowledged Mode}
\newacronym{amc}{AMC}{Adaptive Modulation and Coding}
\newacronym{aqm}{AQM}{Active Queue Management}
\newacronym{awgn}{AGWN}{Additive White Gaussian Noise}
\newacronym{balia}{BALIA}{Balanced Link Adaptation}
\newacronym{bdp}{BDP}{Bandwidth-Delay Product}
\newacronym{bf}{BF}{Beamforming}
\newacronym{cc}{CC}{Congestion Control}
\newacronym{cdf}{CDF}{Cumulative Distribution Function}
\newacronym{cn}{CN}{Core Network}
\newacronym{cqi}{CQI}{Channel Quality Information}
\newacronym{cp}{CP}{Control Plane}
\newacronym{csirs}{CSI-RS}{Channel State Information - Reference Signal}
\newacronym{dc}{DC}{Dual Connectivity}
\newacronym{dce}{DCE}{Direct Code Execution}
\newacronym{dci}{DCI}{Downlink Control Information}
\newacronym{dl}{DL}{Downlink}
\newacronym{dmr}{DMR}{Deadline Miss Ratio}
\newacronym{dmrs}{DMRS}{DeModulation Reference Signal}
\newacronym{e2e}{E2E}{end-to-end}
\newacronym{ecn}{ECN}{Explicit Congestion Notification}
\newacronym{edf}{EDF}{Earliest Deadline First}
\newacronym{enb}{eNB}{evolved Node Base}
\newacronym{epc}{EPC}{Evolved Packet Core}
\newacronym{es}{ES}{Edge Server}
\newacronym{fdma}{FDMA}{Frequency Division Multiple Access}
\newacronym{fdd}{FDD}{Frequency Division Duplexing}
\newacronym[firstplural=Radio Access Technologies (RATs)]{rat}{RAT}{Radio Access Technology}
\newacronym{fs}{FS}{Fast Switching}
\newacronym{ftp}{FTP}{File Transfer Protocol}
\newacronym{gnb}{gNB}{Next Generation Node Base}
\newacronym{harq}{HARQ}{Hybrid Automatic Repeat reQuest}
\newacronym{hetnet}{HetNet}{Heterogeneous Network}
\newacronym{hh}{HH}{Hard Handover}
\newacronym{hol}{HOL}{Head-of-Line}
\newacronym{ia}{IA}{Initial Access}
\newacronym{ieee}{IEEE}{Institute of Electrical and Electronics Engineers}
\newacronym{imt}{IMT}{International Mobile Telecommunication}
\newacronym{iot}{IoT}{Internet of Things}
\newacronym{ldpc}{LDPC}{Low-Density Parity Check}
\newacronym{los}{LOS}{Line-of-Sight}
\newacronym{lte}{LTE}{Long Term Evolution}
\newacronym{m2m}{M2M}{Machine to Machine}
\newacronym{mac}{MAC}{Medium Access Control}
\newacronym{mc}{MC}{Multi-Connectivity}
\newacronym{mcs}{MCS}{Modulation and Coding Scheme}
\newacronym{mec}{MEC}{Mobile Edge Cloud}
\newacronym{mi}{MI}{Mutual Information}
\newacronym{mimo}{MIMO}{Multiple Input, Multiple Output}
\newacronym{mmwave}{mmWave}{millimeter wave}
\newacronym{mptcp}{MPTCP}{Multipath TCP}
\newacronym{mr}{MR}{Maximum Rate}
\newacronym{mss}{MSS}{Maximum Segment Size}
\newacronym{mtd}{MTD}{Machine-Type Device}
\newacronym{mtu}{MTU}{Maximum Transmission Unit}
\newacronym{nfv}{NFV}{Network Function Virtualization}
\newacronym{nlos}{NLOS}{Non-Line-of-Sight}
\newacronym{nlosv}{NLOSv}{Vehicle Non-Line-of-Sight}
\newacronym{nr}{NR}{New Radio}
\newacronym{ofdm}{OFDM}{Orthogonal Frequency Division Multiplexing}
\newacronym{pdcch}{PDCCH}{Physical Downlonk Control Channel}
\newacronym{pdcp}{PDCP}{Packet Data Convergence Protocol}
\newacronym{pdsch}{PDSCH}{Physical Downlink Shared Channel}
\newacronym{pdu}{PDU}{Packet Data Unit}
\newacronym{pf}{PF}{Proportional Fair}
\newacronym{pgw}{PGW}{Packet Gateway}
\newacronym{phy}{PHY}{Physical}
\newacronym{pbch}{PBCH}{Physical Broadcast Channel}
\newacronym[plural=\gls{mme}s,firstplural=Mobility Management Entities (MMEs)]{mme}{MME}{Mobility Management Entity}
\newacronym{prb}{PRB}{Physical Resource Block}
\newacronym{pss}{PSS}{Primary Synchronization Signal}
\newacronym{pscch}{PSCCH}{Physical Sidelink Control Channel}
\newacronym{pucch}{PUCCH}{Physical Uplink Control Channel}
\newacronym{pusch}{PUSCH}{Physical Uplink Shared Channel}
\newacronym{rach}{RACH}{Random Access Channel}
\newacronym{ran}{RAN}{Radio Access Network}
\newacronym{red}{RED}{Random Early Detection}
\newacronym{rf}{RF}{Radio Frequency}
\newacronym{rlc}{RLC}{Radio Link Control}
\newacronym{rlf}{RLF}{Radio Link Failure}
\newacronym{rrc}{RRC}{Radio Resource Control}
\newacronym{rrm}{RRM}{Radio Resource Management}
\newacronym{rr}{RR}{Round Robin}
\newacronym{rs}{RS}{Remote Server}
\newacronym{rsrp}{RSRP}{Reference Signal Received Power}
\newacronym{rss}{RSS}{Received Signal Strength}
\newacronym{rtt}{RTT}{Round Trip Time}
\newacronym{rw}{RW}{Receive Window}
\newacronym{rx}{RX}{Receiver}
\newacronym{sa}{SA}{standalone}
\newacronym{sack}{SACK}{Selective Acknowledgment}
\newacronym{sap}{SAP}{Service Access Point}
\newacronym{sc}{SC}{Single Carrier}
\newacronym{sch}{SCH}{Secondary Cell Handover}
\newacronym{scoot}{SCOOT}{Split Cycle Offset Optimization Technique}
\newacronym{sdma}{SDMA}{Spatial Division Multiple Access}
\newacronym{sinr}{SINR}{Signal to Interference plus Noise Ratio}
\newacronym{sl}{SL}{Sidelink}
\newacronym{sm}{SM}{Saturation Mode}
\newacronym{snr}{SNR}{Signal-to-Noise-Ratio}
\newacronym{son}{SON}{Self-Organizing Network}
\newacronym{ss}{SS}{Synchronization Signal}
\newacronym{srs}{SRS}{Sounding Reference Signal}
\newacronym{sss}{SSS}{Secondary Synchronization Signal}
\newacronym{tb}{TB}{Transport Block}
\newacronym{tcp}{TCP}{Transmission Control Protocol}
\newacronym{tdd}{TDD}{Time Division Duplexing}
\newacronym{tdma}{TDMA}{Time Division Multiple Access}
\newacronym{tfl}{TfL}{Transport for London}
\newacronym{tm}{TM}{Transparent Mode}
\newacronym{trp}{TRP}{Transmitter Receiver Pair}
\newacronym{tti}{TTI}{Transmission Time Interval}
\newacronym{ttt}{TTT}{Time-to-Trigger}
\newacronym{tx}{TX}{Transmitter}
\newacronym{ue}{UE}{User Equipment}
\newacronym{ul}{UL}{Uplink}
\newacronym{uml}{UML}{Unified Modeling Language}
\newacronym{um}{UM}{Unacknowledged Mode}
\newacronym{utc}{UTC}{Urban Traffic Control}
\newacronym{vm}{VM}{Virtual Machine}
\newacronym{rsrq}{RSRQ}{Reference Signal Received Quality}
\newacronym{rssi}{RSSI}{Received Signal Strength Indicator}
\newacronym{crs}{CRS}{Cell Reference Signal}
\newacronym{nsa}{NSA}{Non Stand Alone}
\newacronym{mrdc}{MR-DC}{Multi \gls{rat} \gls{dc}}
\newacronym{endc}{EN-DC}{E-UTRAN-\gls{nr} \gls{dc}}
\newacronym{5gc}{5GC}{5G Core}
\newacronym{si}{SI}{Study Item}
\newacronym{iab}{IAB}{Integrated Access and Backhaul}
\newacronym{wf}{WF}{Wired-first}
\newacronym{hqf}{HQF}{Highest-quality-first}
\newacronym{pa}{PA}{Position-aware}
\newacronym{mlr}{MLR}{Maximum-local-rate}
\newacronym{wbf}{WBF}{Wired Bias Function}
\newacronym{mib}{MIB}{Master Information Block}
\newacronym{sib}{SIB}{Secondary Information Block}
\newacronym{rnti}{RNTI}{Radio Network Temporary Identifier}
\newacronym{dft}{DFT}{Discrete Fourier Transform}
\newacronym{kpi}{KPI}{Key Performance Indicator}
\newacronym{ppp}{PPP}{Poisson Point Process}
\newacronym{v2v}{V2V}{Vehicle-to-Vehicle}
\newacronym{wave}{WAVE}{Wireless Access in Vehicular Environments}
\newacronym{udp}{UDP}{User Datagram Protocol}
\newacronym{upa}{UPA}{Uniform Planar Array}
\newacronym{fec}{FEC}{Forward Error Correction}
\newacronym{v2x}{V2X}{Vehicle-To-Everything}
\newacronym{psfch}{PSFCH}{Physical Sidelink Feedback Channel}
\newacronym{pssch}{PSSCH}{Physical Sidelink Shared Channel}
\newacronym{csma}{CSMA}{Carrier Sense Multiple Access}
\newacronym{v2n}{V2N}{Vehicle-to-Network}
\newacronym{wlan}{WLAN}{Wireless Local Area Network}
\newacronym{cav}{CAV}{Connected and Autonomous Vehicle}
\newacronym{v2i}{V2I}{Vehicle-to-Infrastructure}
\newacronym{d2d}{D2D}{Device-to-Device}
\newacronym{c-its}{C-ITS}{Connected Intelligent Transportation System}
\newacronym{fr2}{FR2}{Frequency Range 2}
\newacronym{bs}{BS}{Base Station}
\newacronym{sdu}{SDU}{Service Data Unit}
\newacronym{csi}{CSI}{Channel State Information}
\newacronym{scs}{SCS}{Subcarrier Spacing}
\newacronym{sumo}{SUMO}{Simulation of Urban MObility}
\newacronym{prr}{PRR}{Packet Reception Ratio}
\newacronym{edca}{EDCA}{Enhanced Distribution Channel Access}

\tikzstyle{startstop} = [rectangle, rounded corners, minimum width=2cm, minimum height=0.5cm,text centered, draw=black]
\tikzstyle{io} = [trapezium, trapezium left angle=70, trapezium right angle=110, minimum width=3cm, minimum height=1cm, text centered, draw=black]
\tikzstyle{process} = [rectangle, minimum width=2cm, minimum height=0.5cm, text centered, draw=black, alignb=center]
\tikzstyle{decision} = [ellipse, minimum width=2cm, minimum height=1cm, text centered, draw=black]
\tikzstyle{arrow} = [thick,<->,>=stealth]
\tikzstyle{line} = [thick,>=stealth]
\tikzstyle{darrow} = [thick,<->,>=stealth,dashed]
\tikzstyle{sarrow} = [thick,->,>=stealth]
\tikzstyle{larrow} = [line width=0.1mm,dashdotted,->,>=stealth]

\makeatletter
\def\grd@save@target#1{%
  \def\grd@target{#1}}
\def\grd@save@start#1{%
  \def\grd@start{#1}}
\tikzset{
  grid with coordinates/.style={
    to path={%
      \pgfextra{%
        \edef\grd@@target{(\tikztotarget)}%
        \tikz@scan@one@point\grd@save@target\grd@@target\relax
        \edef\grd@@start{(\tikztostart)}%
        \tikz@scan@one@point\grd@save@start\grd@@start\relax
        \draw[minor help lines] (\tikztostart) grid (\tikztotarget);
        \draw[major help lines] (\tikztostart) grid (\tikztotarget);
        \grd@start
        \pgfmathsetmacro{\grd@xa}{\the\pgf@x/1cm}
        \pgfmathsetmacro{\grd@ya}{\the\pgf@y/1cm}
        \grd@target
        \pgfmathsetmacro{\grd@xb}{\the\pgf@x/1cm}
        \pgfmathsetmacro{\grd@yb}{\the\pgf@y/1cm}
        \pgfmathsetmacro{\grd@xc}{\grd@xa + \pgfkeysvalueof{/tikz/grid with coordinates/major step x}}
        \pgfmathsetmacro{\grd@yc}{\grd@ya + \pgfkeysvalueof{/tikz/grid with coordinates/major step y}}
        \foreach \x in {\grd@xa,\grd@xc,...,\grd@xb}
        \node[anchor=north] at (\x,\grd@ya) {\pgfmathprintnumber{\x}};
        \foreach \y in {\grd@ya,\grd@yc,...,\grd@yb}
        \node[anchor=east] at (\grd@xa,\y) {\pgfmathprintnumber{\y}};
      }
    }
  },
  minor help lines/.style={
    help lines,
    gray,
    line cap =round,
    xstep=\pgfkeysvalueof{/tikz/grid with coordinates/minor step x},
    ystep=\pgfkeysvalueof{/tikz/grid with coordinates/minor step y}
  },
  major help lines/.style={
    help lines,
    line cap =round,
    line width=\pgfkeysvalueof{/tikz/grid with coordinates/major line width},
    xstep=\pgfkeysvalueof{/tikz/grid with coordinates/major step x},
    ystep=\pgfkeysvalueof{/tikz/grid with coordinates/major step y}
  },
  grid with coordinates/.cd,
  minor step x/.initial=.5,
  minor step y/.initial=.2,
  major step x/.initial=1,
  major step y/.initial=1,
  major line width/.initial=1pt,
}
\makeatother

\makeglossaries
\DeclareUnicodeCharacter{2212}{-} 

\newcommand{\millicar}[0]{MilliCar}

\begin{document}

\title{NR V2X Communications at Millimeter Waves: \\ An End-to-End Performance Evaluation} 

\author{\IEEEauthorblockN{Tommaso Zugno$^*$, Matteo Drago$^*$, Marco Giordani$^*$, Michele Polese$^{\circ}$, Michele Zorzi$^*$}\\
  \IEEEauthorblockA{$^*$Department of Information Engineering, University of Padova, Italy\\email: \texttt{\{name.surname\}@dei.unipd.it}\\
  $^{\circ}$Institute for the Wireless Internet of Things, Northeastern University, Boston, MA\\email: \texttt{m.polese@northeastern.edu}}
  \thanks{This work was partially supported by NIST through Award No. 70NANB17H166.}
}

\flushbottom
\setlength{\parskip}{0ex plus0.1ex}

\maketitle
\thispagestyle{empty}

\glsunset{nr}

\begin{abstract}
3GPP NR V2X represents the new 3GPP standard for next-generation vehicular systems which, among other innovations, supports vehicle-to-vehicle (V2V) operations in the millimeter wave (mmWave) spectrum  to address the communication requirements of future intelligent automotive networks.
While mmWaves will enable massive data rates and low latency, the propagation characteristics at very high frequencies become very challenging,
thereby calling for accurate performance evaluations as a means to properly assess the performance of such systems.
Along these lines, in this paper MilliCar, the new ns-3 module based on the latest NR V2X specifications, is used to provide an end-to-end performance evaluation of  mmWave V2V networks. We investigate the impact of different propagation scenarios and system parameters, including the inter-vehicle distance,  the adopted frame numerology, and the modulation and coding scheme, and provide guidelines towards the most promising V2V deployment configurations.

\end{abstract}

\begin{picture}(0,0)(10,-380)
\put(0,0){
\put(0,0){\footnotesize This paper has been submitted to IEEE Globecom 2020. Copyright may change without notice.}}
\end{picture}


\section{Introduction}
The rapid evolution towards \gls{5g} wireless networks will accelerate the adoption of solutions for \glspl{c-its} to deliver improved traffic safety and efficiency through autonomous driving~\cite{lu2014connected}.
These systems, whose  market estimates  are in the order of 7 trillions USD, promise to make  the number of road accidents drop by as much as 90\%, while carbon emissions will reduce by more than 60\%. The hands-free driving environment of \glspl{c-its} can also reduce drivers' stress and tedium, as well as increase their productivity. \glspl{c-its} could save over 2.7 billion unproductive hours annually in the US in work commutes, according to some estimates~\cite{clements2017economic}.

When fully commercialized, \glspl{c-its} will support several use cases whose requirements will likely exceed the capacity of current communication technologies for vehicular networks~\cite{performance2018giordani,giordani2018feasibility}.
For example, for cooperative perception services, where vehicles exchange processed sensor data to improve the coverage and accuracy of environmental perceptions, the data rate requirements can reach up to approximately 1 Gbps for high-quality uncompressed  images. For advanced safety applications, instead, latency must be very small (i.e., less than 100 ms for high degree of automation) to ensure prompt reactions to unpredictable events~\cite{3GPP_22186}.

\subsection{Towards Millimeter Wave Vehicular Networks} 
\label{sub:millimeter_wave_vehicular_networks}


Going forward into the \gls{5g} era, different standardization activities are currently being promoted as a means to overcome current technology limitations~\cite{zugno2019towards}.
The IEEE 802.11bd standard~\cite{TGbd.general}, for instance, will  target future vehicular service requirements through  new modulation mechanisms,  the introduction of midambles to improve  channel estimation in fast-fading channels, and alternative OFDM numerologies.
Similarly, the 3GPP has agreed on a roadmap to support \glspl{c-its} as part of the NR V2X specifications for Rel. 16~\cite{3gpp.38.885}.
New developments include  sidelink and network architecture improvements, a flexible numerology, and a new resource allocation scheme where sidelink resources can be scheduled autonomously by the vehicles (i.e., mode 2).
In particular, both standards will  support operations at \gls{mmwave} frequencies, up to 71 GHz, where the large available bandwidth can theoretically enable connections with data rates in the order of multiple gigabit per second~\cite{giordani2017millimeter}.

Despite these promises, however,  communication at \glspl{mmwave} in the vehicular environment  introduces several challenges that need to be addressed to ensure robustness and reliability to the end users. First, signals propagating in the  \gls{mmwave} spectrum suffer from severe path and penetration loss, thereby preventing long-range transmissions. Second, directional communications, which are typically established to increase the link budget through beamforming, require precise alignment of the transmitter and the receiver beams and imply increased control overhead.
Also, even though directionality generally guarantees spatial isolation, the metal of vehicles may act as a reflector for the \gls{mmwave} signals, thus resulting in  strong interference from close-by vehicles~\cite{petrov2018impact}.
Finally, the inherently time-varying nature of the \gls{mmwave} channel may prevent long-lived connectivity, with serious impact on the whole protocol stack~\cite{giordani2019lte,zhang2019will}.

Such challenging radio conditions are further exacerbated considering the dynamic topology of the vehicular networks, in particular in the \gls{v2v} scenario for which the direct applicability of the \gls{mmwave} technology is still not clear and has become a research focus in the area of intelligent automotive systems.
As off-the-shelf experimental testbeds operating at \glspl{mmwave} are nowadays very expensive, simulations play a key role in the process of evaluating the performance  of the different design solutions for mmWave vehicular systems.
This is why we recently released \millicar{}\footnote{Available at \url{https://github.com/signetlabdei/millicar}}, the first open-source ns-3 module for V2V mmWave networks~\cite{drago2020millicar}. Compared to the most common  simulators for vehicular networks, e.g., Veins~\cite{sommer2011bidirectionally}, CARLA~\cite{dosovitskiy2017carla} or V2X Simulation Runtime Infrastructure (VSimRTI)~\cite{schunemann2008novel}, \millicar{} enables full-stack end-to-end simulations of the \gls{v2v} network and models the mmWave channel as well as a complete TCP/IP protocol stack and mobility of vehicles.

\subsection{Contributions} 
\label{sub:contributions}


Along these lines, in this paper we target the following two objectives.
First, we validate the main functionalities of the \millicar{} module through an extensive simulation campaign.
We consider a scenario in which two vehicles, one behind the other, are deployed at fixed distance and speed in the same lane, and a scenario in which multiple groups of vehicles  share the same wireless channel on the same road.
Second, we investigate the impact of several system parameters on the end-to-end network performance. More specifically, we examine the effect of the inter-vehicle distance, the propagation scenario, the selected numerology, the \gls{mcs}, and the \gls{rlc} reordering timer on two metrics, i.e., the average communication delay and the \gls{prr}, which are an indication of the robustness of the connection.
We believe that our performance analysis will help stimulate more research on the design and evaluation of mmWave vehicular networks, as well as guide standardization decisions towards the most  promising architectural configuration(s) for V2V deployments.

The rest of this paper is organized as follows.
In Sec.~\ref{sec:millicar}, we describe the features of the \millicar{} module based on~\cite{zugno2020implementation}. In Sec.~\ref{sec:perfeval}, we introduce the simulation scenarios and discuss our main performance results, while in Sec.~\ref{sec:concl} we conclude our work with suggestions for future research.

\section{The MilliCar ns-3 Module}
\label{sec:millicar}

\begin{figure}[t]
  \centering
  \includegraphics[width=\columnwidth]{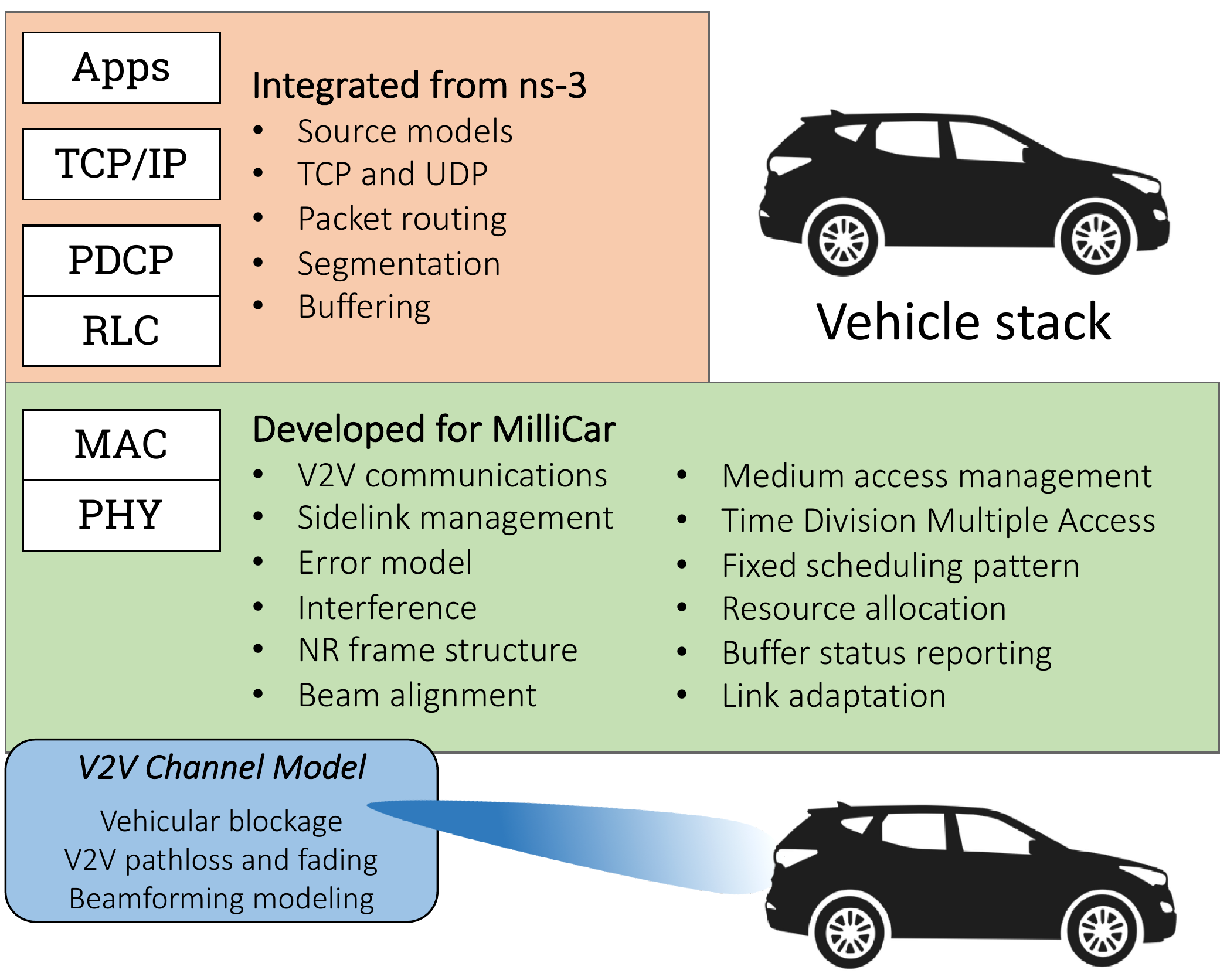}
  \caption{Structure and features of the \millicar{} module.}
  \label{fig:millicar}
\end{figure}

To address the need for a realistic performance evaluation of \gls{v2v} networks at mmWaves, we developed \millicar{}~\cite{drago2020millicar}, an open source \gls{v2v} module for the popular network simulator ns-3~\cite{henderson2008network}. Fig.~\ref{fig:millicar} highlights the main features of the module, which includes the implementation of the 3GPP channel model for \gls{v2v} communications, with propagation, fading and beamformed operations, and physical and \gls{mac} layers compliant with the frame structure of 3GPP NR V2X. Besides, the integration with ns-3 makes it possible to study complex, end-to-end scenarios, with several kinds of applications for the vehicles, the TCP/IP stack, the higher layers of the cellular stack from \gls{lte}~\cite{baldo2011lte}, and realistic mobility patterns, also with the possibility of integrating \gls{sumo} traces~\cite{kuehlmorgen2018simulation}.

The channel model that is implemented in the simulator (and that will be used in the evaluations of this paper) follows the 3GPP report in~\cite{3gpp.37.885}, which specifies a \gls{v2v} model for different scenarios (namely, urban and highway). Notably, the channel can have different conditions, i.e., the link between two devices can be in \gls{los}, if no obstruction is present, \gls{nlosv}, if there is blockage of the \gls{los} path due to other vehicles in the same street, or \gls{nlos}, if the blockage is caused by large environmental obstacles like buildings. Notice that~\cite{3gpp.37.885} defines probabilistic transitions only between \gls{los} and \gls{nlosv}, while the \gls{nlos} condition is given by the geometry of the scenario. Therefore, we also give the possibility to use the model in~\cite{boban2016evolution} for a fully statistical channel characterization. Pathloss and fading follow the equations in~\cite{3gpp.37.885}, which vary according to the specific channel condition. Beamforming is modeled with a \gls{dft}-based approach that points to the \gls{los} direction between two vehicles, and the possibility of changing the number of antenna elements in the vehicles. Interference is accounted for as well, and the error model that maps the \gls{sinr} into an error probability for the transport blocks of the physical layer is taken from~\cite{mezzavilla2018end}.

\begin{figure*}[b]
\begin{subfigure}[t]{0.48\textwidth}
  \setlength{\abovecaptionskip}{-0.25cm}
  \centering
  \setlength\fwidth{\columnwidth}
  \setlength\fheight{.6\columnwidth}
  \input{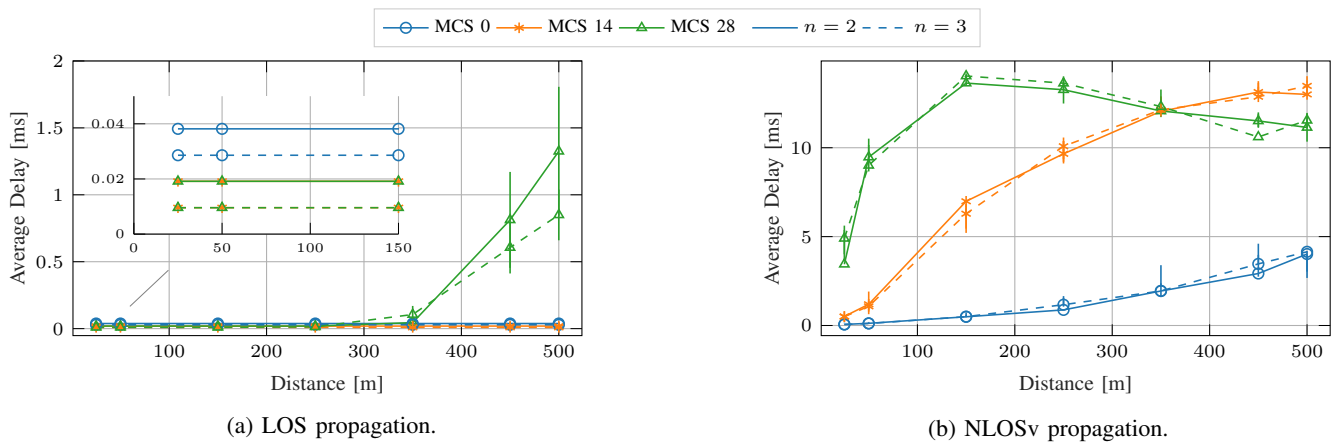}
  \caption{LOS propagation.}
  \label{fig:highwayLOSdelay}
\end{subfigure}%
\hfill%
\begin{subfigure}[t]{0.48\textwidth}
  \centering
  \setlength\fwidth{\columnwidth}
  \setlength\fheight{0.6\columnwidth}
\begin{tikzpicture}

  \definecolor{color0}{rgb}{0.12156862745098,0.466666666666667,0.705882352941177}
  \definecolor{color1}{rgb}{1,0.498039215686275,0.0549019607843137}
  \definecolor{color2}{rgb}{0.172549019607843,0.627450980392157,0.172549019607843}
  \pgfplotsset{every tick label/.append style={font=\scriptsize}}
\pgfplotsset{every tick label/.append style={font=\scriptsize}}

\begin{axis}[
  width=0.951\fwidth,
  height=\fheight,
  at={(0\fwidth,0\fheight)},
legend cell align={left},
legend style={fill opacity=0.8, draw opacity=1, text opacity=1, at={(0.91,0.5)}, anchor=east, draw=white!80!black},
tick pos=both,
x grid style={white!69.0196078431373!black},
xlabel={Distance [m]},
xmajorgrids,
xmin=1.25, xmax=523.75,
xtick style={color=black},
y grid style={white!69.0196078431373!black},
ylabel={Average Delay [ms]},
ylabel style={yshift=-0.15cm, font=\footnotesize\color{white!15!black}},
xlabel style={font=\footnotesize\color{white!15!black}},
ymajorgrids,
ymin=-0.568604452601095, ymax=14.8813142677713,
ytick style={color=black}
]
\path [draw=color0, semithick]
(axis cs:25,0.133836698364679)
--(axis cs:25,0.148374924475194);

\path [draw=color0, semithick]
(axis cs:50,0.159990517653124)
--(axis cs:50,0.199808121851608);

\path [draw=color0, semithick]
(axis cs:150,0.468604741450117)
--(axis cs:150,0.737232769341229);

\path [draw=color0, semithick]
(axis cs:250,1.09836812997971)
--(axis cs:250,1.55469603473202);

\path [draw=color0, semithick]
(axis cs:350,2.18047121390204)
--(axis cs:350,3.392600752247);

\path [draw=color0, semithick]
(axis cs:450,2.74642297677493)
--(axis cs:450,3.9720981981377);

\path [draw=color0, semithick]
(axis cs:500,2.67522277367135)
--(axis cs:500,3.98671020866031);

\path [draw=color1, semithick]
(axis cs:25,0.292167146051932)
--(axis cs:25,0.451390821200136);

\path [draw=color1, semithick]
(axis cs:50,1.04215471040214)
--(axis cs:50,1.90925916116127);

\path [draw=color1, semithick]
(axis cs:150,5.20907734912864)
--(axis cs:150,6.74784533363967);

\path [draw=color1, semithick]
(axis cs:250,9.114831516054)
--(axis cs:250,10.372779625122);

\path [draw=color1, semithick]
(axis cs:350,11.929807651902)
--(axis cs:350,12.8675207648592);

\path [draw=color1, semithick]
(axis cs:450,13.0398517314048)
--(axis cs:450,13.7463422040465);

\path [draw=color1, semithick]
(axis cs:500,13.6594195147867)
--(axis cs:500,14.0247196163149);

\path [draw=color2, semithick]
(axis cs:25,4.00700481740834)
--(axis cs:25,5.61864164812145);

\path [draw=color2, semithick]
(axis cs:50,8.66488707107042)
--(axis cs:50,10.153606679984);

\path [draw=color2, semithick]
(axis cs:150,13.9310128119532)
--(axis cs:150,14.1790452350271);

\path [draw=color2, semithick]
(axis cs:250,12.487938287821)
--(axis cs:250,14.0136466393117);

\path [draw=color2, semithick]
(axis cs:350,11.7215441455079)
--(axis cs:350,12.822712901217);

\path [draw=color2, semithick]
(axis cs:450,11.1221908948364)
--(axis cs:450,11.9888031786302);

\path [draw=color2, semithick]
(axis cs:500,10.3424693063674)
--(axis cs:500,11.9146447513123);

\path [draw=color0, semithick]
(axis cs:25,0.133664580143105)
--(axis cs:25,0.153408755722745);

\path [draw=color0, semithick]
(axis cs:50,0.154893914191352)
--(axis cs:50,0.183298065580652);

\path [draw=color0, semithick]
(axis cs:150,0.466806571023589)
--(axis cs:150,0.600309089529767);

\path [draw=color0, semithick]
(axis cs:250,1.05793003244604)
--(axis cs:250,1.65360336610461);

\path [draw=color0, semithick]
(axis cs:350,1.59612008046374)
--(axis cs:350,2.3728238175296);

\path [draw=color0, semithick]
(axis cs:450,3.42611595985903)
--(axis cs:450,4.60128337008329);

\path [draw=color0, semithick]
(axis cs:500,3.02756007309742)
--(axis cs:500,4.18871581384865);

\path [draw=color1, semithick]
(axis cs:25,0.239780734372523)
--(axis cs:25,0.425874050643924);

\path [draw=color1, semithick]
(axis cs:50,0.635480672275363)
--(axis cs:50,1.12427633462935);

\path [draw=color1, semithick]
(axis cs:150,5.44876867213677)
--(axis cs:150,6.8383296652622);

\path [draw=color1, semithick]
(axis cs:250,9.20904897883712)
--(axis cs:250,10.5743984246056);

\path [draw=color1, semithick]
(axis cs:350,12.0867077775908)
--(axis cs:350,12.8951482768296);

\path [draw=color1, semithick]
(axis cs:450,13.0833934483852)
--(axis cs:450,13.6528424265648);

\path [draw=color1, semithick]
(axis cs:500,12.9483987289319)
--(axis cs:500,13.5679354687392);

\path [draw=color2, semithick]
(axis cs:25,3.67383837340213)
--(axis cs:25,5.37408269084637);

\path [draw=color2, semithick]
(axis cs:50,9.26727266341362)
--(axis cs:50,10.511738849386);

\path [draw=color2, semithick]
(axis cs:150,13.4861848800175)
--(axis cs:150,14.085401514195);

\path [draw=color2, semithick]
(axis cs:250,13.2995308638104)
--(axis cs:250,13.9227244115094);

\path [draw=color2, semithick]
(axis cs:350,12.5108821550894)
--(axis cs:350,13.2749151975672);

\path [draw=color2, semithick]
(axis cs:450,11.1704415227438)
--(axis cs:450,11.9126982992508);

\path [draw=color2, semithick]
(axis cs:500,11.1066408368889)
--(axis cs:500,11.9593595296889);

\addplot [semithick, color0, mark=o, mark options={solid}, forget plot]
table {%
25 0.0607589804583876
50 0.122296783074508
150 0.490147449869896
250 0.87747185542526
350 1.94108493781529
450 2.92401273926954
500 4.019628101541
};
\addplot [semithick, color1, mark=asterisk, mark options={solid}, forget plot]
table {%
25 0.509344527402512
50 1.19286964163224
150 6.98706559008543
250 9.65915727069882
350 12.0507869758708
450 13.1315630506342
500 13.0006662443987
};
\addplot [semithick, color2, mark=triangle, mark options={solid}, forget plot]
table {%
25 3.44629577751079
50 9.47126070323556
150 13.635282007608
250 13.2638245511802
350 12.0682649567315
450 11.5082694419259
500 11.1417063735923
};
\addplot [semithick, color0, dashed, mark=o, mark options={solid}, forget plot]
table {%
25 0.0577548393522478
50 0.112395794517053
150 0.496446660336258
250 1.16947648939117
350 1.94479283594748
450 3.46611130602217
500 4.14688312152165
};
\addplot [semithick, color1, dashed, mark=asterisk, mark options={solid}, forget plot]
table {%
25 0.496167317738235
50 1.07873109444752
150 6.28737770932714
250 10.0824840743881
350 12.1365196571601
450 12.8675850976871
500 13.4788432007521
};
\addplot [semithick, color2, dashed, mark=triangle, mark options={solid}, forget plot]
table {%
25 4.90094027424101
50 9.0080666542162
150 14.0356701298683
250 13.630615350179
350 12.3160664054138
450 10.6050061662422
500 11.5437528525709
};
\end{axis}

\end{tikzpicture}
  \caption{NLOSv propagation.}
  \label{fig:highwayNLOSvdelay}
\end{subfigure}
\caption{Performance comparison of different numerologies ($n$) and \glspl{mcs}, for a highway scenario.}
\label{fig:highwaydelaycomp}
\end{figure*}

The physical layer implementation models the frame structure for NR V2X~\cite{3gpp.38.885}, which inherits two of the \gls{ofdm} numerologies of NR, i.e., numerology 2, with a subcarrier spacing of 60 kHz and 4 slots of 14 symbols for each subframe (1 ms), and numerology 3, with 120 kHz as subcarrier spacing and 8 slots. The \gls{mac} layer is \gls{tdma}-based, with transmissions from different vehicles assigned to different slots, according to a pre-established scheduling pattern. Additionally, \millicar{} adopts the \gls{amc} scheme of the ns-3 mmWave module~\cite{mezzavilla2018end}, with the possibility of configuring a fixed \gls{mcs} when feedback on the channel condition is not available. Finally, the custom implementation of the physical and \gls{mac} layers is connected to the \gls{rlc} and \gls{pdcp} layers of the \gls{lte} module, which handle the forwarding of packets to the TCP/IP stack, and buffering and segmentation to match the transport block size at the \gls{mac} layer.

\begin{table}[t!]
  \caption{Simulation parameters.}
  \label{table:params}
  \centering
  \begin{tabular}{lll}
    \toprule
     & \textbf{Scenario A} & \textbf{Scenario B}\\
    \midrule
    Distance & $\left[25, \dots, 500\right]$ m& $40$ m\\
    Speed & $20$ m/s & $20$ m/s\\
    Propagation scenario & [Urban , Highway]& Highway\\
    Antenna size & $4\times4$& $\left[1\times1, 2\times2, 4\times4 \right]$\\
    Bandwidth & $100$ MHz & 100 MHz\\
    Carrier frequency &$28$ GHz & $28$ GHz\\
    Numerology & $\left[2, 3\right]$& 3\\
    MCS & $\left[0, 14, 28 \right]$& $\left[0, 28 \right]$\\
    \gls{rlc} mode& Unacknowledged& Unacknowledged\\
    \gls{rlc} reordering timer & $\left[1, \dots, 100 \right]$ ms & $10$ ms\\
    \gls{rlc} buffer size & 512 kBytes & 512 kBytes\\
    UDP source rate & $800$ kbps & $\left[10, 50, 100 \right]$ Mbps\\
    \bottomrule
  \end{tabular}
\end{table}

\section{Performance Evaluation}
\label{sec:perfeval}
In this section, we present the results of the performance evaluation conducted with the MilliCar module.
We developed two simulation scenarios, i.e., Scenario~A,  described in Sec.~\ref{sec:scenarioA}, and Scenario~B, described in Sec.~\ref{sec:scenarioB}.
In particular, the former has been designed to analyze the impact of different system parameters, namely the \gls{mcs}, the inter-vehicle distance, the \gls{rlc} configuration, and the selected numerology, on the end-to-end communication performance, and to compare the system behavior in different propagation scenarios.
The latter, instead, considers the presence of multiple groups of vehicles travelling on the same road and sharing the wireless channel, and evaluates the performance achieved with different modulation and coding schemes and antenna settings.
For each scenario, we carried out a simulation campaign and computed some metrics of interest by averaging the results of multiple independent runs.
Table~\ref{table:params} summarizes the parameters used in our simulations.

\subsection{Impact of Numerology, \gls{mcs}, and RLC Parameters}
\label{sec:scenarioA}

In Scenario A, two vehicles proceed one in front of the other at a constant speed of $20$ m/s, keeping the same distance during the whole simulation. One vehicle, i.e., the server, generates packets of $100$~Bytes at a fixed rate of $800$~kbps, and sends them to the other vehicle, i.e., the client, using a  \gls{udp} application. 

First, we studied the performance of the end-to-end delay and \gls{prr} at increasing values of the inter-vehicle distance, focusing on the 3GPP Highway scenario. We assessed these metrics for \gls{los} and \gls{nlosv} channel conditions and, for a more detailed insight, we compared the results obtained using different numerologies (i.e., $n=2$ and $n=3$) and \glspl{mcs}.
In Figure~\ref{fig:highwayLOSdelay} we show that, in the \gls{los} regime, at lower distances numerology $3$ guarantees the lowest average delay. This is motivated by the fact that, as described in Sec.~\ref{sec:millicar}, for this numerology the subframe in divided into 8 slots (with respect to 4 slots for numerology 2), resulting in shorter \gls{ofdm} symbols, to fit the same subframe duration.
On the other hand, it is more difficult to observe a difference between the different numerologies if we consider the \gls{nlosv} channel condition, as illustrated in Figure~\ref{fig:highwayNLOSvdelay}, which generally results in a significantly higher end-to-end delay compared to the \gls{los} case. In  \gls{nlosv}, in fact, some packets can be lost due to a bad channel state. In addition, as the receiving \gls{rlc} entity implements a reordering procedure for all the received \glspl{pdu}, it has to wait for missing packets in the receiving window until the reordering timer expires: this may increase the packet delay regardless of the numerology that is selected.
It should also be mentioned that,  when using \gls{mcs} 28 in \gls{los},  the average delay grows remarkably if we increase the inter-vehicle distance above $250$ m as a result of degraded channel conditions, as shown in Figure~\ref{fig:highwayLOSdelay}. However,  at such long distances, in a real scenario the path would likely be obstructed by other vehicles and, in this case, we expect that the delay will evolve as shown in Figure~\ref{fig:highwayNLOSvdelay} for the NLOSv regime.
Moreover, it can be noticed that for \gls{mcs}~28 in NLOSv, the average delay shows a decreasing behavior when considering distances above 150~m. This is a consequence of the high packet loss rate experienced at such distances, which results in less congested buffers for the remaining packets.
As a side note, our results also confirm that better resilience is offered by \gls{mcs} 0, which guarantees a delay as low as around $1.3$ ms, even at $250$ m in \gls{nlosv}.

\begin{figure*}
\begin{subfigure}[b]{0.3\textwidth}
  \setlength{\abovecaptionskip}{-0.25cm}
	\centering
	\setlength\fwidth{1.2\columnwidth}
	\setlength\fheight{.85\columnwidth}
\begin{tikzpicture}

  \definecolor{color0}{rgb}{0.12156862745098,0.466666666666667,0.705882352941177}
  \definecolor{color2}{rgb}{1,0.498039215686275,0.0549019607843137}
  \definecolor{color1}{rgb}{0.172549019607843,0.627450980392157,0.172549019607843}
\pgfplotsset{every tick label/.append style={font=\scriptsize}}

\begin{axis}[
  width=0.951\fwidth,
  height=\fheight,
  at={(0\fwidth,0\fheight)},
legend cell align={left},
legend style={fill opacity=0.8, draw opacity=1, text opacity=1, at={(0.03,0.03)}, anchor=south west, draw=white!80!black},
tick align=inside,
tick pos=both,
x grid style={white!69.0196078431373!black},
xlabel={Distance [m]},
xmajorgrids,
xmin=1.25, xmax=523.75,
xtick style={color=black},
y grid style={white!69.0196078431373!black},
ylabel={PRR},
ylabel style={yshift=-0.15cm, font=\footnotesize\color{white!15!black}},
xlabel style={font=\footnotesize\color{white!15!black}},
ymajorgrids,
ymin=0.680736385047418, ymax=1.01533944705161,
ytick style={color=black},
ytick={0.65,0.7,0.75,0.8,0.85,0.9,0.95,1,1.05},
yticklabels={0.65,0.70,0.75,0.80,0.85,0.90,0.95,1.00,1.05}
]
\path [draw=color0, semithick]
(axis cs:25,1)
--(axis cs:25,1);

\path [draw=color0, semithick]
(axis cs:50,1)
--(axis cs:50,1);

\path [draw=color0, semithick]
(axis cs:150,1)
--(axis cs:150,1);

\path [draw=color0, semithick]
(axis cs:250,1)
--(axis cs:250,1);

\path [draw=color0, semithick]
(axis cs:350,1)
--(axis cs:350,1);

\path [draw=color0, semithick]
(axis cs:450,1)
--(axis cs:450,1);

\path [draw=color0, semithick]
(axis cs:500,1)
--(axis cs:500,1);

\path [draw=color1, semithick]
(axis cs:25,1)
--(axis cs:25,1);

\path [draw=color1, semithick]
(axis cs:50,1)
--(axis cs:50,1);

\path [draw=color1, semithick]
(axis cs:150,1)
--(axis cs:150,1);

\path [draw=color1, semithick]
(axis cs:250,0.999362236404402)
--(axis cs:250,1.0000565670144);

\path [draw=color1, semithick]
(axis cs:350,0.976534481104651)
--(axis cs:350,0.992525347955178);

\path [draw=color1, semithick]
(axis cs:450,0.822490230094425)
--(axis cs:450,0.909543957939763);

\path [draw=color1, semithick]
(axis cs:500,0.695945615138517)
--(axis cs:500,0.814037290844388);

\path [draw=color0, semithick]
(axis cs:25,1)
--(axis cs:25,1);

\path [draw=color0, semithick]
(axis cs:50,1)
--(axis cs:50,1);

\path [draw=color0, semithick]
(axis cs:150,1)
--(axis cs:150,1);

\path [draw=color0, semithick]
(axis cs:250,1)
--(axis cs:250,1);

\path [draw=color0, semithick]
(axis cs:350,1)
--(axis cs:350,1);

\path [draw=color0, semithick]
(axis cs:450,1)
--(axis cs:450,1);

\path [draw=color0, semithick]
(axis cs:500,1)
--(axis cs:500,1);

\path [draw=color1, semithick]
(axis cs:25,1)
--(axis cs:25,1);

\path [draw=color1, semithick]
(axis cs:50,1)
--(axis cs:50,1);

\path [draw=color1, semithick]
(axis cs:150,1)
--(axis cs:150,1);

\path [draw=color1, semithick]
(axis cs:250,0.999356962526667)
--(axis cs:250,1.00013021696051);

\path [draw=color1, semithick]
(axis cs:350,0.999828523340445)
--(axis cs:350,1.0000347245228);

\path [draw=color1, semithick]
(axis cs:450,0.999378452355852)
--(axis cs:450,1.00004035106295);

\path [draw=color1, semithick]
(axis cs:500,0.996610725627249)
--(axis cs:500,0.999081582065059);

\addplot [semithick, color0, mark=o, mark options={solid}, forget plot]
table {%
25 1
50 1
150 1
250 1
350 1
450 1
500 1
};
\addplot [semithick, color1, mark=triangle, mark options={solid}, forget plot]
table {%
25 1
50 1
150 1
250 0.999709401709402
350 0.984529914529915
450 0.866017094017094
500 0.754991452991453
};
\addplot [semithick, color0, dashed, mark=o, mark options={solid}, forget plot]
table {%
25 1
50 1
150 1
250 1
350 1
450 1
500 1
};
\addplot [semithick, color1, dashed, mark=triangle, mark options={solid}, forget plot]
table {%
25 1
50 1
150 1
250 0.99974358974359
350 0.999931623931624
450 0.999709401709402
500 0.997846153846154
};
\end{axis}

\end{tikzpicture}
	\caption{LOS propagation}
  \label{fig:prrLOS}
\end{subfigure}
\hfill%
\begin{subfigure}[b]{0.3\textwidth}
  \setlength{\abovecaptionskip}{-0.25cm}
	\centering
	\setlength\fwidth{1.2\columnwidth}
	\setlength\fheight{.85\columnwidth}
\begin{tikzpicture}

  \definecolor{color0}{rgb}{0.12156862745098,0.466666666666667,0.705882352941177}
  \definecolor{color2}{rgb}{1,0.498039215686275,0.0549019607843137}
  \definecolor{color1}{rgb}{0.172549019607843,0.627450980392157,0.172549019607843}
\pgfplotsset{every tick label/.append style={font=\scriptsize}}

\begin{axis}[
  width=0.951\fwidth,
  height=\fheight,
  at={(0\fwidth,0\fheight)},
legend cell align={left},
legend style={fill opacity=0.8, draw opacity=1, text opacity=1, at={(0.03,0.03)}, anchor=south west, draw=white!80!black},
tick align=inside,
tick pos=both,
x grid style={white!69.0196078431373!black},
xlabel={Distance [m]},
xmajorgrids,
xmin=1.25, xmax=523.75,
xtick style={color=black},
y grid style={white!69.0196078431373!black},
ylabel={PRR},
ylabel style={yshift=-0.15cm, font=\footnotesize\color{white!15!black}},
xlabel style={font=\footnotesize\color{white!15!black}},
ymajorgrids,
ymin=-0.04314189629657, ymax=1.04979084324967,
ytick style={color=black},
ytick={-0.2,0,0.2,0.4,0.6,0.8,1,1.2},
yticklabels={−0.2,0.0,0.2,0.4,0.6,0.8,1.0,1.2}
]
\path [draw=color0, semithick]
(axis cs:25,0.999277453242938)
--(axis cs:25,0.999696905731421);

\path [draw=color0, semithick]
(axis cs:50,0.998178427981339)
--(axis cs:50,0.998847213044302);

\path [draw=color0, semithick]
(axis cs:150,0.990339447256805)
--(axis cs:150,0.993592176674819);

\path [draw=color0, semithick]
(axis cs:250,0.978592817374866)
--(axis cs:250,0.984552481770433);

\path [draw=color0, semithick]
(axis cs:350,0.965350119003804)
--(axis cs:350,0.974137060483376);

\path [draw=color0, semithick]
(axis cs:450,0.916088104565391)
--(axis cs:450,0.960082835605549);

\path [draw=color0, semithick]
(axis cs:500,0.916425435269309)
--(axis cs:500,0.94706174421787);

\path [draw=color1, semithick]
(axis cs:25,0.889880464344034)
--(axis cs:25,0.934837484373915);

\path [draw=color1, semithick]
(axis cs:50,0.798139481565737)
--(axis cs:50,0.845860518434263);

\path [draw=color1, semithick]
(axis cs:150,0.350128066472223)
--(axis cs:150,0.439444583100427);

\path [draw=color1, semithick]
(axis cs:250,0.142128582281546)
--(axis cs:250,0.218401921697234);

\path [draw=color1, semithick]
(axis cs:350,0.0443826820755921)
--(axis cs:350,0.0866518006830286);

\path [draw=color1, semithick]
(axis cs:450,0.00653686459189544)
--(axis cs:450,0.0251002162759547);

\path [draw=color1, semithick]
(axis cs:500,0.00859916556412557)
--(axis cs:500,0.0256665687016087);

\path [draw=color0, semithick]
(axis cs:25,0.996742618493496)
--(axis cs:25,1.0001120823612);

\path [draw=color0, semithick]
(axis cs:50,0.998747084668084)
--(axis cs:50,0.999577701656702);

\path [draw=color0, semithick]
(axis cs:150,0.994874712456708)
--(axis cs:150,0.996851783269787);

\path [draw=color0, semithick]
(axis cs:250,0.990022082431468)
--(axis cs:250,0.994285609876224);

\path [draw=color0, semithick]
(axis cs:350,0.973349120546173)
--(axis cs:350,0.988633785436733);

\path [draw=color0, semithick]
(axis cs:450,0.977098253814904)
--(axis cs:450,0.987106874390224);

\path [draw=color0, semithick]
(axis cs:500,0.974255258756362)
--(axis cs:500,0.981847305346202);

\path [draw=color1, semithick]
(axis cs:25,0.934419509076168)
--(axis cs:25,0.954879636222977);

\path [draw=color1, semithick]
(axis cs:50,0.836773722407778)
--(axis cs:50,0.888012602378547);

\path [draw=color1, semithick]
(axis cs:150,0.463336200297521)
--(axis cs:150,0.570509953548633);

\path [draw=color1, semithick]
(axis cs:250,0.349438714794959)
--(axis cs:250,0.453467268111024);

\path [draw=color1, semithick]
(axis cs:350,0.216460668806755)
--(axis cs:350,0.300462408116322);

\path [draw=color1, semithick]
(axis cs:450,0.115500984476482)
--(axis cs:450,0.173517583162775);

\path [draw=color1, semithick]
(axis cs:500,0.0691847503372355)
--(axis cs:500,0.102168034808653);

\addplot [semithick, color0, mark=o, mark options={solid}, forget plot]
table {%
25 0.999487179487179
50 0.99851282051282
150 0.991965811965812
250 0.98157264957265
350 0.96974358974359
450 0.93808547008547
500 0.93174358974359
};
\addplot [semithick, color1, mark=triangle, mark options={solid}, forget plot]
table {%
25 0.912358974358974
50 0.822
150 0.394786324786325
250 0.18026525198939
350 0.0655172413793103
450 0.015818540433925
500 0.0171328671328671
};
\addplot [semithick, color0, dashed, mark=o, mark options={solid}, forget plot]
table {%
25 0.99842735042735
50 0.999162393162393
150 0.995863247863248
250 0.992153846153846
350 0.980991452991453
450 0.982102564102564
500 0.978051282051282
};
\addplot [semithick, color1, dashed, mark=triangle, mark options={solid}, forget plot]
table {%
25 0.944649572649573
50 0.862393162393163
150 0.516923076923077
250 0.401452991452991
350 0.258461538461538
450 0.144509283819629
500 0.0856763925729443
};
\end{axis}

%
%
%

\begin{axis}[
width=0.951\fwidth,
height=\fheight,
at={(0\fwidth,0\fheight)},
legend cell align={left},
legend style={font=\scriptsize,at={(0.5,1.05)}, anchor=south, draw=white!80.0!black},
tick align=inside,
tick pos=left,
x grid style={white!69.01960784313725!black},
xmajorgrids,
xmin=0, xmax=840,
xtick style={color=black},
y grid style={white!69.01960784313725!black},
ymajorgrids,
ymin=0, ymax=810,
ytick style={color=black},
hide y axis,
hide x axis,
legend columns=6,
]

\addplot [solid, semithick, mark=o, color0]
table [row sep=crcr] {%
-1 -1\\
-2 -2\\
};
\addlegendentry{MCS 0}

\addplot [solid, semithick, mark=triangle, color1]
table [row sep=crcr] {%
-1 -1\\
-2 -2\\
};
\addlegendentry{MCS 28}
 \addlegendimage{empty legend}\addlegendentry{}

\addplot [solid, semithick, color0]
table [row sep=crcr] {%
-1 -1\\
-2 -2\\
};
\addlegendentry{Highway}

\addplot [dashed, semithick, color0]
table [row sep=crcr] {%
-1 -1\\
-2 -2\\
};
\addlegendentry{Urban}

\end{axis}

\end{tikzpicture}
	\caption{NLOSv propagation}
  \label{fig:prrNLOSv}
\end{subfigure}
\hfill%
\begin{subfigure}[b]{0.3\textwidth}
  \setlength{\abovecaptionskip}{-0.25cm}
	\centering
	\setlength\fwidth{1.2\columnwidth}
	\setlength\fheight{.85\columnwidth}
\begin{tikzpicture}

  \definecolor{color0}{rgb}{0.12156862745098,0.466666666666667,0.705882352941177}
  \definecolor{color2}{rgb}{1,0.498039215686275,0.0549019607843137}
  \definecolor{color1}{rgb}{0.172549019607843,0.627450980392157,0.172549019607843}
\pgfplotsset{every tick label/.append style={font=\scriptsize}}

\begin{axis}[
  width=0.951\fwidth,
  height=\fheight,
  at={(0\fwidth,0\fheight)},
legend cell align={left},
legend style={fill opacity=0.8, draw opacity=1, text opacity=1, draw=white!80!black},
tick align=inside,
tick pos=both,
x grid style={white!69.0196078431373!black},
xlabel={Distance [m]},
xmajorgrids,
xmin=1.25, xmax=523.75,
xtick style={color=black},
y grid style={white!69.0196078431373!black},
ylabel={PRR},
ylabel style={yshift=-0.15cm, font=\footnotesize\color{white!15!black}},
xlabel style={font=\footnotesize\color{white!15!black}},
ymajorgrids,
ymin=-0.0491778833594889, ymax=1.04985985712509,
ytick style={color=black},
ytick={-0.2,0,0.2,0.4,0.6,0.8,1,1.2},
yticklabels={−0.2,0.0,0.2,0.4,0.6,0.8,1.0,1.2}
]
\path [draw=color0, semithick]
(axis cs:25,0.999686147395773)
--(axis cs:25,0.999903596193971);

\path [draw=color0, semithick]
(axis cs:50,0.99413705053822)
--(axis cs:50,0.998102265701096);

\path [draw=color0, semithick]
(axis cs:150,0.882227257699152)
--(axis cs:150,0.928097528625634);

\path [draw=color0, semithick]
(axis cs:250,0.709992992609552)
--(axis cs:250,0.800365981749422);

\path [draw=color0, semithick]
(axis cs:350,0.443557149750551)
--(axis cs:350,0.577945944855991);

\path [draw=color0, semithick]
(axis cs:450,0.281743718581922)
--(axis cs:450,0.409728266223396);

\path [draw=color0, semithick]
(axis cs:500,0.195818151245156)
--(axis cs:500,0.316134511476738);

\path [draw=color1, semithick]
(axis cs:25,0.824170564907663)
--(axis cs:25,0.890256785519688);

\path [draw=color1, semithick]
(axis cs:50,0.322114233655899)
--(axis cs:50,0.465931743355596);

\path [draw=color1, semithick]
(axis cs:150,0.000884322357500862)
--(axis cs:150,0.00209936062618212);

\path [draw=color0, semithick]
(axis cs:25,0.999063043677745)
--(axis cs:25,0.999603622988921);

\path [draw=color0, semithick]
(axis cs:50,0.987517827928846)
--(axis cs:50,0.994636018225);

\path [draw=color0, semithick]
(axis cs:150,0.75343972067188)
--(axis cs:150,0.845329510097351);

\path [draw=color0, semithick]
(axis cs:250,0.447246702834091)
--(axis cs:250,0.58752252793514);

\path [draw=color0, semithick]
(axis cs:350,0.191474721232309)
--(axis cs:350,0.312957513199925);

\path [draw=color0, semithick]
(axis cs:450,0.0567255198244906)
--(axis cs:450,0.143981818813883);

\path [draw=color0, semithick]
(axis cs:500,0.0323017805931527)
--(axis cs:500,0.094291626000254);

\path [draw=color1, semithick]
(axis cs:25,0.649061562471005)
--(axis cs:25,0.744716215306773);

\path [draw=color1, semithick]
(axis cs:50,0.0735640013460057)
--(axis cs:50,0.148853581071577);

\path [draw=color1, semithick]
(axis cs:150,0.000778377571628268)
--(axis cs:150,0.00101649422324353);

\addplot [semithick, color0, mark=o, mark options={solid}, forget plot]
table {%
25 0.999794871794872
50 0.996119658119658
150 0.905162393162393
250 0.755179487179487
350 0.510751547303271
450 0.345735992402659
500 0.255976331360947
};
\addplot [semithick, color1, mark=triangle, mark options={solid}, forget plot]
table {%
25 0.857213675213675
50 0.394022988505747
150 0.00149184149184149
250 0.0
350 0.0
450 0.0
500 0.0
};
\addplot [semithick, color0, dashed, mark=o, mark options={solid}, forget plot]
table {%
25 0.999333333333333
50 0.991076923076923
150 0.799384615384615
250 0.517384615384615
350 0.252216117216117
450 0.100353669319187
500 0.0632967032967033
};
\addplot [semithick, color1, dashed, mark=triangle, mark options={solid}, forget plot]
table {%
25 0.696888888888889
50 0.111208791208791
150 0.000897435897435897
250 0.0
350 0.0
450 0.0
500 0.0
};
\end{axis}

\end{tikzpicture}
	\caption{NLOS propagation}
  \label{fig:prrNLOS}
\end{subfigure}
\caption{\gls{prr} for different channel conditions and propagation scenarios, numerology  $n=3$, and packet size $100$ Bytes.}
\label{fig:prr}
\end{figure*}
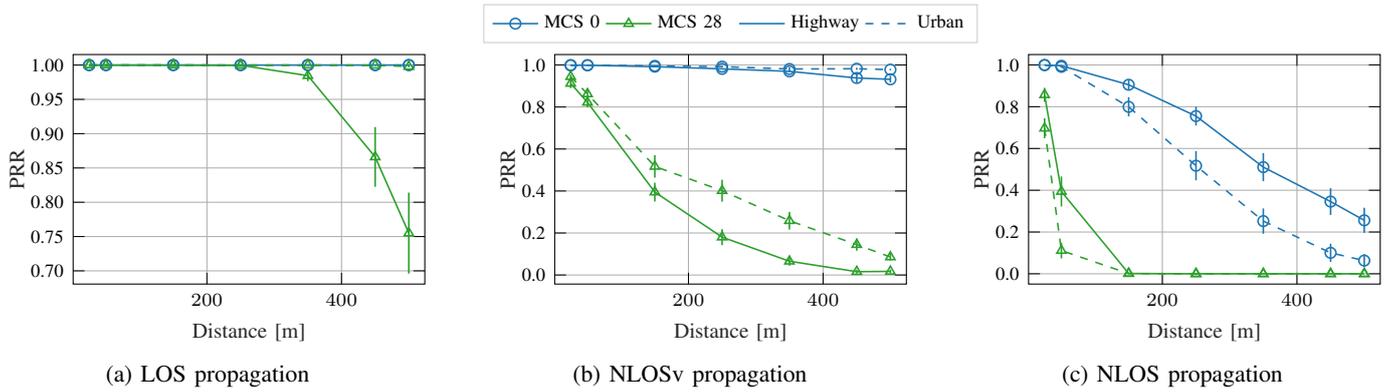

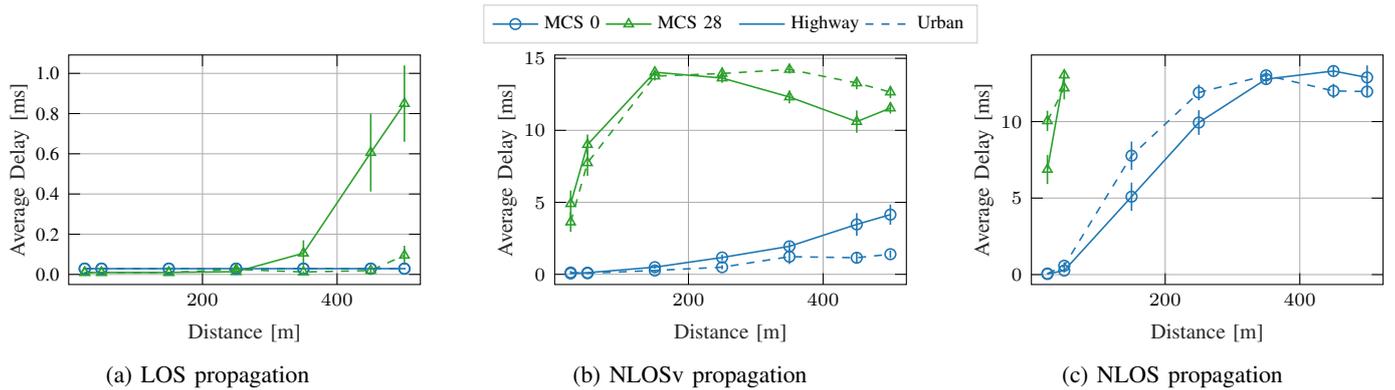
\begin{figure*}
\begin{subfigure}[t]{0.3\textwidth}
  \setlength{\abovecaptionskip}{-0.25cm}
	\centering
	\setlength\fwidth{1.2\columnwidth}
	\setlength\fheight{.85\columnwidth}
\begin{tikzpicture}

  \definecolor{color0}{rgb}{0.12156862745098,0.466666666666667,0.705882352941177}
  \definecolor{color2}{rgb}{1,0.498039215686275,0.0549019607843137}
  \definecolor{color1}{rgb}{0.172549019607843,0.627450980392157,0.172549019607843}
\pgfplotsset{every tick label/.append style={font=\scriptsize}}

\begin{axis}[
  width=0.951\fwidth,
  height=\fheight,
  at={(0\fwidth,0\fheight)},
legend cell align={left},
legend style={fill opacity=0.8, draw opacity=1, text opacity=1, at={(0.03,0.97)}, anchor=north west, draw=white!80!black},
tick align=inside,
tick pos=both,
x grid style={white!69.0196078431373!black},
xlabel={Distance [m]},
xmajorgrids,
xmin=1.25, xmax=523.75,
xtick style={color=black},
y grid style={white!69.0196078431373!black},
ylabel={Average Delay [ms]},
ylabel style={yshift=-0.15cm, font=\footnotesize\color{white!15!black}},
xlabel style={font=\footnotesize\color{white!15!black}},
ymajorgrids,
ymin=-0.0502202708912408, ymax=1.09146465368505,
ytick style={color=black},
ytick={-0.2,0,0.2,0.4,0.6,0.8,1,1.2},
yticklabels={−0.2,0.0,0.2,0.4,0.6,0.8,1.0,1.2}
]
\path [draw=color0, semithick]
(axis cs:25,0.0285974358974107)
--(axis cs:25,0.0285974358974107);

\path [draw=color0, semithick]
(axis cs:50,0.0285974358974107)
--(axis cs:50,0.0285974358974107);

\path [draw=color0, semithick]
(axis cs:150,0.0285974358974107)
--(axis cs:150,0.0285974358974107);

\path [draw=color0, semithick]
(axis cs:250,0.0285974358974107)
--(axis cs:250,0.0285974358974107);

\path [draw=color0, semithick]
(axis cs:350,0.0285974358974107)
--(axis cs:350,0.0285974358974107);

\path [draw=color0, semithick]
(axis cs:450,0.0285974358974107)
--(axis cs:450,0.0285974358974107);

\path [draw=color0, semithick]
(axis cs:500,0.0285974358974107)
--(axis cs:500,0.0285974358974107);

\path [draw=color1, semithick]
(axis cs:25,0.0095743589743485)
--(axis cs:25,0.0095743589743485);

\path [draw=color1, semithick]
(axis cs:50,0.0095743589743485)
--(axis cs:50,0.0095743589743485);

\path [draw=color1, semithick]
(axis cs:150,0.0095743589743485)
--(axis cs:150,0.0095743589743485);

\path [draw=color1, semithick]
(axis cs:250,0.00907745623628118)
--(axis cs:250,0.0175154121685831);

\path [draw=color1, semithick]
(axis cs:350,0.0413505727437831)
--(axis cs:350,0.169625483282199);

\path [draw=color1, semithick]
(axis cs:450,0.41223585793157)
--(axis cs:450,0.799342261453221);

\path [draw=color1, semithick]
(axis cs:500,0.660074437956764)
--(axis cs:500,1.03956988438612);

\path [draw=color0, semithick]
(axis cs:25,0.0285974358974107)
--(axis cs:25,0.0285974358974107);

\path [draw=color0, semithick]
(axis cs:50,0.0285974358974107)
--(axis cs:50,0.0285974358974107);

\path [draw=color0, semithick]
(axis cs:150,0.0285974358974107)
--(axis cs:150,0.0285974358974107);

\path [draw=color0, semithick]
(axis cs:250,0.0285974358974107)
--(axis cs:250,0.0285974358974107);

\path [draw=color0, semithick]
(axis cs:350,0.0285974358974107)
--(axis cs:350,0.0285974358974107);

\path [draw=color0, semithick]
(axis cs:450,0.0285974358974107)
--(axis cs:450,0.0285974358974107);

\path [draw=color0, semithick]
(axis cs:500,0.0285974358974107)
--(axis cs:500,0.0285974358974107);

\path [draw=color1, semithick]
(axis cs:25,0.0095743589743485)
--(axis cs:25,0.0095743589743485);

\path [draw=color1, semithick]
(axis cs:50,0.0095743589743485)
--(axis cs:50,0.0095743589743485);

\path [draw=color1, semithick]
(axis cs:150,0.0095743589743485)
--(axis cs:150,0.0095743589743485);

\path [draw=color1, semithick]
(axis cs:250,0.00167449840768132)
--(axis cs:250,0.0485854552131816);

\path [draw=color1, semithick]
(axis cs:350,0.00842610911770113)
--(axis cs:350,0.0152446471124587);

\path [draw=color1, semithick]
(axis cs:450,0.00864639419467439)
--(axis cs:450,0.0302224125579633);

\path [draw=color1, semithick]
(axis cs:500,0.04726020420005)
--(axis cs:500,0.142382994513116);

\addplot [semithick, color0, mark=o, mark options={solid}, forget plot]
table {%
25 0.0285974358974107
50 0.0285974358974107
150 0.0285974358974107
250 0.0285974358974107
350 0.0285974358974107
450 0.0285974358974107
500 0.0285974358974107
};
\addplot [semithick, color1, mark=triangle, mark options={solid}, forget plot]
table {%
25 0.0095743589743485
50 0.0095743589743485
150 0.0095743589743485
250 0.0132964342024322
350 0.105488028012991
450 0.605789059692395
500 0.849822161171444
};
\addplot [semithick, color0, dashed, mark=o, mark options={solid}, forget plot]
table {%
25 0.0285974358974107
50 0.0285974358974107
150 0.0285974358974107
250 0.0285974358974107
350 0.0285974358974107
450 0.0285974358974107
500 0.0285974358974107
};
\addplot [semithick, color1, dashed, mark=triangle, mark options={solid}, forget plot]
table {%
25 0.0095743589743485
50 0.0095743589743485
150 0.0095743589743485
250 0.0251299768104315
350 0.0118353781150799
450 0.0194344033763189
500 0.0948215993565829
};
\end{axis}

\end{tikzpicture}
	\caption{LOS propagation}
  \label{fig:delayLOS}
\end{subfigure}
\hfill%
\begin{subfigure}[t]{0.3\textwidth}
  \setlength{\abovecaptionskip}{-0.25cm}
	\centering
	\setlength\fwidth{1.2\columnwidth}
	\setlength\fheight{.85\columnwidth}
\begin{tikzpicture}

  \definecolor{color0}{rgb}{0.12156862745098,0.466666666666667,0.705882352941177}
  \definecolor{color2}{rgb}{1,0.498039215686275,0.0549019607843137}
  \definecolor{color1}{rgb}{0.172549019607843,0.627450980392157,0.172549019607843}
\pgfplotsset{every tick label/.append style={font=\scriptsize}}

\begin{axis}[
  width=0.951\fwidth,
  height=\fheight,
  at={(0\fwidth,0\fheight)},
legend cell align={left},
legend style={fill opacity=0.8, draw opacity=1, text opacity=1, at={(0.91,0.5)}, anchor=east, draw=white!80!black},
tick align=inside,
tick pos=both,
x grid style={white!69.0196078431373!black},
xlabel={Distance [m]},
xmajorgrids,
xmin=1.25, xmax=523.75,
xtick style={color=black},
y grid style={white!69.0196078431373!black},
ylabel={Average Delay [ms]},
ylabel style={yshift=-0.15cm, font=\footnotesize\color{white!15!black}},
xlabel style={font=\footnotesize\color{white!15!black}},
ymajorgrids,
ymin=-0.70667117669839, ymax=15.2317052999407,
ytick style={color=black}
]
\path [draw=color0, semithick]
(axis cs:25,0.0459256223136721)
--(axis cs:25,0.0695840563908234);

\path [draw=color0, semithick]
(axis cs:50,0.0927967981270081)
--(axis cs:50,0.131994790907097);

\path [draw=color0, semithick]
(axis cs:150,0.396992042183395)
--(axis cs:150,0.59590127848912);

\path [draw=color0, semithick]
(axis cs:250,0.968153247981503)
--(axis cs:250,1.37079973080084);

\path [draw=color0, semithick]
(axis cs:350,1.64574913288354)
--(axis cs:350,2.24383653901143);

\path [draw=color0, semithick]
(axis cs:450,2.68042155834696)
--(axis cs:450,4.25180105369737);

\path [draw=color0, semithick]
(axis cs:500,3.44574271774954)
--(axis cs:500,4.84802352529377);

\path [draw=color1, semithick]
(axis cs:25,3.98190959154713)
--(axis cs:25,5.81997095693489);

\path [draw=color1, semithick]
(axis cs:50,8.3072324515184)
--(axis cs:50,9.70890085691399);

\path [draw=color1, semithick]
(axis cs:150,13.8815995288524)
--(axis cs:150,14.1897407308843);

\path [draw=color1, semithick]
(axis cs:250,13.2980455927233)
--(axis cs:250,13.9631851076348);

\path [draw=color1, semithick]
(axis cs:350,11.8695727010161)
--(axis cs:350,12.7625601098116);

\path [draw=color1, semithick]
(axis cs:450,9.82949198433938)
--(axis cs:450,11.380520348145);

\path [draw=color1, semithick]
(axis cs:500,11.1634411298868)
--(axis cs:500,11.9240645752551);

\path [draw=color0, semithick]
(axis cs:25,0.0178004813306614)
--(axis cs:25,0.236472699475288);

\path [draw=color0, semithick]
(axis cs:50,0.0516303977533992)
--(axis cs:50,0.0989196429940686);

\path [draw=color0, semithick]
(axis cs:150,0.20549148516168)
--(axis cs:150,0.329150048840763);

\path [draw=color0, semithick]
(axis cs:250,0.353248829922053)
--(axis cs:250,0.637791791393845);

\path [draw=color0, semithick]
(axis cs:350,0.716064855000522)
--(axis cs:350,1.73405551973286);

\path [draw=color0, semithick]
(axis cs:450,0.798487926006769)
--(axis cs:450,1.50491780145554);

\path [draw=color0, semithick]
(axis cs:500,1.14137938848293)
--(axis cs:500,1.65237168197893);

\path [draw=color1, semithick]
(axis cs:25,2.95078865833621)
--(axis cs:25,4.30310366848617);

\path [draw=color1, semithick]
(axis cs:50,6.84132568984432)
--(axis cs:50,8.64023120855862);

\path [draw=color1, semithick]
(axis cs:150,13.4618507460955)
--(axis cs:150,14.0668340791599);

\path [draw=color1, semithick]
(axis cs:250,13.527824319697)
--(axis cs:250,14.3668244260356);

\path [draw=color1, semithick]
(axis cs:350,13.9441826025766)
--(axis cs:350,14.5072336419117);

\path [draw=color1, semithick]
(axis cs:450,12.8512006679618)
--(axis cs:450,13.7211041326297);

\path [draw=color1, semithick]
(axis cs:500,12.2328658727617)
--(axis cs:500,13.0786873388287);

\addplot [semithick, color0, mark=o, mark options={solid}, forget plot]
table {%
25 0.0577548393522478
50 0.112395794517053
150 0.496446660336258
250 1.16947648939117
350 1.94479283594748
450 3.46611130602217
500 4.14688312152165
};
\addplot [semithick, color1, mark=triangle, mark options={solid}, forget plot]
table {%
25 4.90094027424101
50 9.0080666542162
150 14.0356701298683
250 13.630615350179
350 12.3160664054138
450 10.6050061662422
500 11.5437528525709
};
\addplot [semithick, color0, dashed, mark=o, mark options={solid}, forget plot]
table {%
25 0.127136590402975
50 0.0752750203737339
150 0.267320767001221
250 0.495520310657949
350 1.22506018736669
450 1.15170286373115
500 1.39687553523093
};
\addplot [semithick, color1, dashed, mark=triangle, mark options={solid}, forget plot]
table {%
25 3.62694616341119
50 7.74077844920147
150 13.7643424126277
250 13.9473243728663
350 14.2257081222442
450 13.2861524002957
500 12.6557766057952
};
\end{axis}

%
%


\begin{axis}[
width=0.951\fwidth,
height=\fheight,
at={(0\fwidth,0\fheight)},
legend cell align={left},
legend style={font=\scriptsize,at={(0.5,1.05)}, anchor=south, draw=white!80.0!black},
tick align=inside,
tick pos=left,
x grid style={white!69.01960784313725!black},
xmajorgrids,
xmin=0, xmax=840,
xtick style={color=black},
y grid style={white!69.01960784313725!black},
ymajorgrids,
ymin=0, ymax=810,
ytick style={color=black},
hide y axis,
hide x axis,
legend columns=6,
]

\addplot [solid, semithick, mark=o, color0]
table [row sep=crcr] {%
-1 -1\\
-2 -2\\
};
\addlegendentry{MCS 0}

\addplot [solid, semithick, mark=triangle, color1]
table [row sep=crcr] {%
-1 -1\\
-2 -2\\
};
\addlegendentry{MCS 28}
 \addlegendimage{empty legend}\addlegendentry{}

\addplot [solid, semithick, color0]
table [row sep=crcr] {%
-1 -1\\
-2 -2\\
};
\addlegendentry{Highway}

\addplot [dashed, semithick, color0]
table [row sep=crcr] {%
-1 -1\\
-2 -2\\
};
\addlegendentry{Urban}

\end{axis}

\end{tikzpicture}
	\caption{NLOSv propagation}
  \label{fig:delayNLOSv}
\end{subfigure}
\hfill%
\begin{subfigure}[t]{0.3\textwidth}
  \setlength{\abovecaptionskip}{-0.25cm}
	\centering
	\setlength\fwidth{1.2\columnwidth}
	\setlength\fheight{.85\columnwidth}
\begin{tikzpicture}

  \definecolor{color0}{rgb}{0.12156862745098,0.466666666666667,0.705882352941177}
  \definecolor{color2}{rgb}{1,0.498039215686275,0.0549019607843137}
  \definecolor{color1}{rgb}{0.172549019607843,0.627450980392157,0.172549019607843}
  \pgfplotsset{every tick label/.append style={font=\scriptsize}}

\begin{axis}[
  width=0.951\fwidth,
  height=\fheight,
  at={(0\fwidth,0\fheight)},
legend cell align={left},
legend style={fill opacity=0.8, draw opacity=1, text opacity=1, at={(0.97,0.03)}, anchor=south east, draw=white!80!black},
tick align=inside,
tick pos=both,
x grid style={white!69.0196078431373!black},
xlabel={Distance [m]},
xmajorgrids,
xmin=1.25, xmax=523.75,
xtick style={color=black},
y grid style={white!69.0196078431373!black},
ylabel={Average Delay [ms]},
ylabel style={yshift=-0.15cm, font=\footnotesize\color{white!15!black}},
xlabel style={font=\footnotesize\color{white!15!black}},
ymajorgrids,
ymin=-0.647976129404893, ymax=14.3533703494107,
ytick style={color=black}
]
\path [draw=color0, semithick]
(axis cs:25,0.0339032559958176)
--(axis cs:25,0.0458769963066944);

\path [draw=color0, semithick]
(axis cs:50,0.134760128551503)
--(axis cs:50,0.394593468795936);

\path [draw=color0, semithick]
(axis cs:150,4.16924109455048)
--(axis cs:150,6.0128123991223);

\path [draw=color0, semithick]
(axis cs:250,9.13398210675274)
--(axis cs:250,10.7425891565907);

\path [draw=color0, semithick]
(axis cs:350,12.3981666358877)
--(axis cs:350,13.1603042419355);

\path [draw=color0, semithick]
(axis cs:450,12.9926748067594)
--(axis cs:450,13.6188886641817);

\path [draw=color0, semithick]
(axis cs:500,12.0998883168047)
--(axis cs:500,13.67149096401);

\path [draw=color1, semithick]
(axis cs:25,5.91992737356938)
--(axis cs:25,7.82253259875691);

\path [draw=color1, semithick]
(axis cs:50,12.7324301451275)
--(axis cs:50,13.3425789962665);


\path [draw=color0, semithick]
(axis cs:25,0.0501669412269106)
--(axis cs:25,0.0832258741693113);

\path [draw=color0, semithick]
(axis cs:50,0.349484042020247)
--(axis cs:50,0.806635656849566);

\path [draw=color0, semithick]
(axis cs:150,6.84838949317762)
--(axis cs:150,8.69533517168633);

\path [draw=color0, semithick]
(axis cs:250,11.3887072603427)
--(axis cs:250,12.4143834839504);

\path [draw=color0, semithick]
(axis cs:350,12.7374626066115)
--(axis cs:350,13.3089366049096);

\path [draw=color0, semithick]
(axis cs:450,11.5387087970137)
--(axis cs:450,12.4709771936219);

\path [draw=color0, semithick]
(axis cs:500,11.5893543624201)
--(axis cs:500,12.3378710509704);

\path [draw=color1, semithick]
(axis cs:25,9.3809185944044)
--(axis cs:25,10.7110301442642);

\path [draw=color1, semithick]
(axis cs:50,11.4509930571072)
--(axis cs:50,12.928097028466);

\addplot [semithick, color0, mark=o, mark options={solid}, forget plot]
table {%
25 0.039890126151256
50 0.26467679867372
150 5.09102674683639
250 9.93828563167172
350 12.7792354389116
450 13.3057817354705
500 12.8856896404074
};
\addplot [semithick, color1, mark=triangle, mark options={solid}, forget plot]
table {%
25 6.87122998616315
50 13.037504570697
250 nan
350 nan
450 nan
500 nan
};
\addplot [semithick, color0, dashed, mark=o, mark options={solid}, forget plot]
table {%
25 0.0666964076981109
50 0.578059849434907
150 7.77186233243197
250 11.9015453721466
350 13.0231996057606
450 12.0048429953178
500 11.9636127066952
};
\addplot [semithick, color1, dashed, mark=triangle, mark options={solid}, forget plot]
table {%
25 10.0459743693343
50 12.1895450427866
250 nan
350 nan
450 nan
500 nan
};
\end{axis}

\end{tikzpicture}
	\caption{NLOS propagation}
  \label{fig:delayNLOS}
\end{subfigure}
\caption{Average delay for different channel conditions and propagation scenarios, numerology $n=3$, and packet size $100$ Bytes.}
\label{fig:delay}
\end{figure*}

In Figures~\ref{fig:prr} and ~\ref{fig:delay} we plot the \gls{prr} and average delay, respectively, as a function of the inter-vehicle distance and the  channel conditions, i.e., urban or highway, for a fixed numerology $n=3$.
In particular, Figures~\ref{fig:prrLOS} and \ref{fig:prrNLOSv} exemplify that better end-to-end performance can be obtained using the urban path loss configuration with \gls{los} and \gls{nlosv} conditions, hence resulting in a lower latency, as represented by Figures~\ref{fig:delayLOS} and \ref{fig:delayNLOSv}. This is motivated by the fact that, in an urban environment, the communication benefits from reflections from walls and/or environmental blockages, which are more likely in street canyons. In this scenario, however, static objects are also more likely to completely block the signal, thus resulting in communication outage. This is demonstrated by Figures~\ref{fig:prrNLOS} and \ref{fig:delayNLOS}, where the trend of the curves is switched and urban propagation results in reduced PRR (up to $-80\%$) and increased latency (up to $+50\%$) compared to highway propagation.

Finally, in Figure~\ref{fig:timer} we study how the average end-to-end delay and the \gls{prr} are affected by  different values of the \gls{rlc} reordering timer. In particular, we can see from Figure~\ref{fig:timerPRR} that higher values of the reordering timer do not significantly affect the reception ratio, which is almost constant inside the confidence intervals regardless of the modulation and coding scheme. Since we are using \gls{rlc} unacknowledged mode and we are not implementing any \gls{harq} techniques at the \gls{mac} layer, lost packets are not retransmitted\footnote{The \gls{prr} is expected to improve with higher values of the reordering timer when retransmissions are used, but this study is left for future work.}, therefore, if there are some missing packets in the receiving window, the reordering timer associated to each packet has to expire before they can be forwarded to the upper layers, which results in an increased  experienced delay, as shown in Figure~\ref{fig:timerdelay}.

\begin{figure*}
\begin{subfigure}[t]{0.48\textwidth}
  \setlength{\abovecaptionskip}{-0.25cm}
  \centering
  \setlength\fwidth{\columnwidth}
  \setlength\fheight{0.6\columnwidth}
\begin{tikzpicture}

  \definecolor{color0}{rgb}{0.12156862745098,0.466666666666667,0.705882352941177}
  \definecolor{color2}{rgb}{1,0.498039215686275,0.0549019607843137}
  \definecolor{color1}{rgb}{0.172549019607843,0.627450980392157,0.172549019607843}
\pgfplotsset{every tick label/.append style={font=\scriptsize}}

\begin{axis}[
  width=0.951\fwidth,
  height=\fheight,
  at={(0\fwidth,0\fheight)},
legend cell align={left},
legend style={fill opacity=0.8, draw opacity=1, text opacity=1, at={(0.5,0.09)}, anchor=south, draw=white!80!black},
tick align=inside,
tick pos=both,
x grid style={white!69.0196078431373!black},
xlabel={RLC Reordering Timer [ms]},
xmajorgrids,
xmin=-3.95, xmax=104.95,
xtick style={color=black},
y grid style={white!69.0196078431373!black},
ylabel={PRR},
ylabel style={yshift=-0.15cm, font=\footnotesize\color{white!15!black}},
xlabel style={font=\footnotesize\color{white!15!black}},
ymajorgrids,
ymin=0.209855176092552, ymax=1.03676591006411,
ytick style={color=black},
ytick={0.2,0.3,0.4,0.5,0.6,0.7,0.8,0.9,1,1.1},
yticklabels={0.2,0.3,0.4,0.5,0.6,0.7,0.8,0.9,1.0,1.1}
]
\path [draw=color0, semithick]
(axis cs:1,0.987817788509227)
--(axis cs:1,0.996035691344253);

\path [draw=color0, semithick]
(axis cs:10,0.988199371277494)
--(axis cs:10,0.994687075608953);

\path [draw=color0, semithick]
(axis cs:40,0.988408565082528)
--(axis cs:40,0.9957965631226);

\path [draw=color0, semithick]
(axis cs:70,0.985731311336082)
--(axis cs:70,0.995689934085163);

\path [draw=color0, semithick]
(axis cs:100,0.986226321283746)
--(axis cs:100,0.993319466262041);

\path [draw=color1, semithick]
(axis cs:1,0.247442027636713)
--(axis cs:1,0.390535994341309);

\path [draw=color1, semithick]
(axis cs:10,0.344852033258827)
--(axis cs:10,0.491895219488425);

\path [draw=color1, semithick]
(axis cs:40,0.416891306403392)
--(axis cs:40,0.526788934923909);

\path [draw=color1, semithick]
(axis cs:70,0.288227535010649)
--(axis cs:70,0.428607629824516);

\path [draw=color1, semithick]
(axis cs:100,0.35307877441882)
--(axis cs:100,0.487565914225869);

\path [draw=color0, semithick]
(axis cs:1,0.995642063095799)
--(axis cs:1,0.997852442398707);

\path [draw=color0, semithick]
(axis cs:10,0.994867205124021)
--(axis cs:10,0.997660267403452);

\path [draw=color0, semithick]
(axis cs:40,0.99519919621882)
--(axis cs:40,0.997211060191437);

\path [draw=color0, semithick]
(axis cs:70,0.990711051370164)
--(axis cs:70,0.999179058519946);

\path [draw=color0, semithick]
(axis cs:100,0.993223390059069)
--(axis cs:100,0.99819052935485);

\path [draw=color1, semithick]
(axis cs:1,0.453338233961117)
--(axis cs:1,0.614471289848407);

\path [draw=color1, semithick]
(axis cs:10,0.466231380508435)
--(axis cs:10,0.619131256854202);

\path [draw=color1, semithick]
(axis cs:40,0.505080274404867)
--(axis cs:40,0.654568077243485);

\path [draw=color1, semithick]
(axis cs:70,0.551666974562972)
--(axis cs:70,0.651028996132999);

\path [draw=color1, semithick]
(axis cs:100,0.460101034754963)
--(axis cs:100,0.607034496380568);

\addplot [semithick, color0, mark=o, mark size=2, mark options={solid}, forget plot]
table {%
1 0.99192673992674
10 0.991443223443223
40 0.992102564102564
70 0.990710622710623
100 0.989772893772894
};
\addplot [semithick, color1, mark=triangle, mark size=2, mark options={solid}, forget plot]
table {%
1 0.318989010989011
10 0.418373626373626
40 0.47184012066365
70 0.358417582417582
100 0.420322344322344
};
\addplot [semithick, color0, dashed, mark=o, mark size=2, mark options={solid}, forget plot]
table {%
1 0.996747252747253
10 0.996263736263736
40 0.996205128205128
70 0.994945054945055
100 0.99570695970696
};
\addplot [semithick, color1, dashed, mark=triangle, mark size=2, mark options={solid}, forget plot]
table {%
1 0.533904761904762
10 0.542681318681319
40 0.579824175824176
70 0.601347985347985
100 0.533567765567766
};
\end{axis}

\begin{axis}[
width=0.951\fwidth,
height=\fheight,
at={(0\fwidth,0\fheight)},
legend cell align={left},
legend style={font=\scriptsize,at={(1.25,1.05)}, anchor=south, draw=white!80.0!black},
tick pos=left,
x grid style={white!69.01960784313725!black},
xmajorgrids,
xmin=0, xmax=840,
xtick style={color=black},
y grid style={white!69.01960784313725!black},
ymajorgrids,
ymin=0, ymax=810,
ytick style={color=black},
hide y axis,
hide x axis,
legend columns=6,
]

\addplot [solid, semithick, mark=o, color0]
table [row sep=crcr] {%
-1 -1\\
-2 -2\\
};
\addlegendentry{MCS 0}

\addplot [solid, semithick, mark=triangle, color1]
table [row sep=crcr] {%
-1 -1\\
-2 -2\\
};
\addlegendentry{MCS 28}

\addlegendimage{empty legend}\addlegendentry{}

\addplot [solid, semithick, color0]
table [row sep=crcr] {%
-1 -1\\
-2 -2\\
};
\addlegendentry{Urban}

\addplot [dashed, semithick, color0]
table [row sep=crcr] {%
-1 -1\\
-2 -2\\
};
\addlegendentry{Highway}

\end{axis}

\end{tikzpicture}
  \caption{PRR}
  \label{fig:timerPRR}
\end{subfigure}
\hfill
\begin{subfigure}[t]{0.48\textwidth}
  \centering
  \setlength\fwidth{\columnwidth}
  \setlength\fheight{.6\columnwidth}
\begin{tikzpicture}

  \definecolor{color0}{rgb}{0.12156862745098,0.466666666666667,0.705882352941177}
  \definecolor{color2}{rgb}{1,0.498039215686275,0.0549019607843137}
  \definecolor{color1}{rgb}{0.172549019607843,0.627450980392157,0.172549019607843}
\pgfplotsset{every tick label/.append style={font=\scriptsize}}

\begin{axis}[
  width=0.951\fwidth,
  height=\fheight,
  at={(0\fwidth,0\fheight)},
legend cell align={left},
legend style={fill opacity=0.8, draw opacity=1, text opacity=1, at={(0.03,0.97)}, anchor=north west, draw=white!80!black},
tick align=inside,
tick pos=both,
x grid style={white!69.0196078431373!black},
xlabel={RLC Reordering Timer [ms]},
xmajorgrids,
xmin=-3.95, xmax=104.95,
xtick style={color=black},
y grid style={white!69.0196078431373!black},
ylabel={Average Delay [ms]},
ylabel style={yshift=-0.15cm, font=\footnotesize\color{white!15!black}},
xlabel style={font=\footnotesize\color{white!15!black}},
ymajorgrids,
ymin=-7.46820948418331, ymax=157.508220463816,
ytick style={color=black}
]
\path [draw=color0, semithick]
(axis cs:1,0.0325299831583474)
--(axis cs:1,0.0407840672521356);

\path [draw=color0, semithick]
(axis cs:10,0.325830862172734)
--(axis cs:10,0.745286427077177);

\path [draw=color0, semithick]
(axis cs:40,4.17270233990184)
--(axis cs:40,10.1928796199224);

\path [draw=color0, semithick]
(axis cs:70,15.4516069107388)
--(axis cs:70,29.4149642215841);

\path [draw=color0, semithick]
(axis cs:100,39.8956748586919)
--(axis cs:100,63.0738526776293);

\path [draw=color1, semithick]
(axis cs:1,0.545155423048835)
--(axis cs:1,0.675358765234351);

\path [draw=color1, semithick]
(axis cs:10,13.4833431695456)
--(axis cs:10,14.1064581604218);

\path [draw=color1, semithick]
(axis cs:40,58.648310155494)
--(axis cs:40,59.432257916818);

\path [draw=color1, semithick]
(axis cs:70,103.784500319479)
--(axis cs:70,104.827406936524);

\path [draw=color1, semithick]
(axis cs:100,149.179126342697)
--(axis cs:100,150.009291829817);

\path [draw=color0, semithick]
(axis cs:1,0.0307191498166841)
--(axis cs:1,0.032931313569885);

\path [draw=color0, semithick]
(axis cs:10,0.158784999904039)
--(axis cs:10,0.320735284586105);

\path [draw=color0, semithick]
(axis cs:40,2.51617262392282)
--(axis cs:40,4.59479782054537);

\path [draw=color0, semithick]
(axis cs:70,5.57046067197345)
--(axis cs:70,17.1691468428075);

\path [draw=color0, semithick]
(axis cs:100,11.8550671302483)
--(axis cs:100,33.0728989088714);

\path [draw=color1, semithick]
(axis cs:1,0.392704264067999)
--(axis cs:1,0.553033799540943);

\path [draw=color1, semithick]
(axis cs:10,13.167287250941)
--(axis cs:10,14.137318634157);

\path [draw=color1, semithick]
(axis cs:40,57.9236791738035)
--(axis cs:40,59.1512199316213);

\path [draw=color1, semithick]
(axis cs:70,102.959361035066)
--(axis cs:70,103.904233137722);

\path [draw=color1, semithick]
(axis cs:100,148.617819976538)
--(axis cs:100,149.598420229391);

\addplot [semithick, color0, mark=o, mark size=2, mark options={solid}, forget plot]
table {%
1 0.0366570252052415
10 0.535558644624956
40 7.1827909799121
70 22.4332855661615
100 51.4847637681606
};
\addplot [semithick, color1, mark=triangle, mark size=2, mark options={solid}, forget plot]
table {%
1 0.610257094141593
10 13.7949006649837
40 59.040284036156
70 104.305953628002
100 149.594209086257
};
\addplot [semithick, color0, dashed, mark=o, mark size=2, mark options={solid}, forget plot]
table {%
1 0.0318252316932846
10 0.239760142245072
40 3.5554852222341
70 11.3698037573905
100 22.4639830195598
};
\addplot [semithick, color1, dashed, mark=triangle, mark size=2, mark options={solid}, forget plot]
table {%
1 0.472869031804471
10 13.652302942549
40 58.5374495527124
70 103.431797086394
100 149.108120102965
};
\end{axis}

\end{tikzpicture}
  \caption{Average delay}
  \label{fig:timerdelay}
\end{subfigure}%
\caption{Performance comparison as a function of the RLC reordering timer, for numerology  $n=3$, NLOSv channel condition, and inter-vehicle distance equal to $150$ m.}
\label{fig:timer}
\end{figure*}
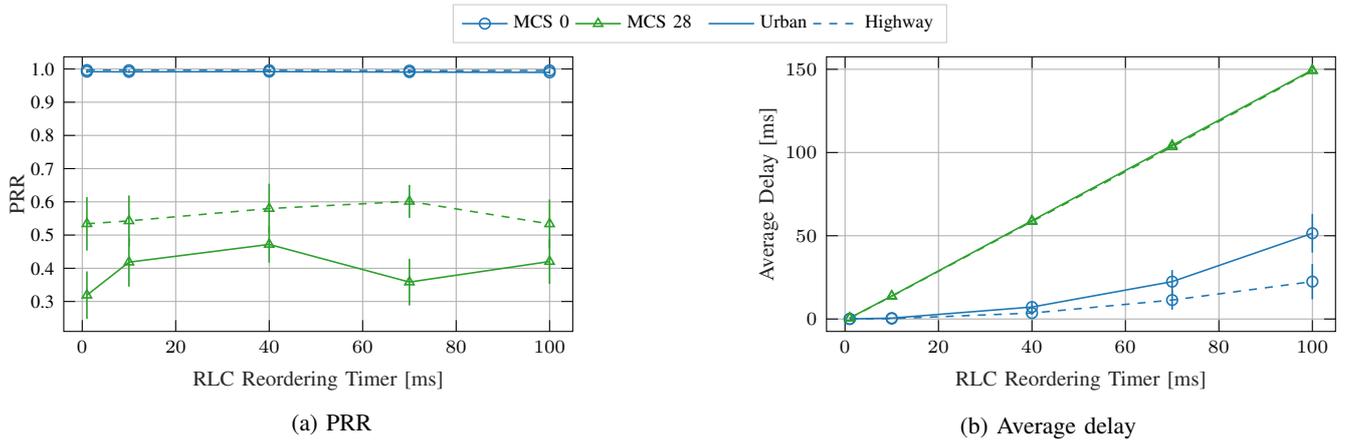

\begin{figure*}
  \begin{subfigure}[t]{0.48\textwidth}
    \setlength{\abovecaptionskip}{-0.25cm}
    \centering
    \setlength\fwidth{\columnwidth}
    \setlength\fheight{.6\columnwidth}
\begin{tikzpicture}

\definecolor{color0}{rgb}{0.12156862745098,0.466666666666667,0.705882352941177}
\definecolor{color1}{rgb}{1,0.498039215686275,0.0549019607843137}
\definecolor{color2}{rgb}{0.172549019607843,0.627450980392157,0.172549019607843}
\pgfplotsset{every tick label/.append style={font=\scriptsize}}

\begin{axis}[
  width=0.951\fwidth,
  height=\fheight,
  at={(0\fwidth,0\fheight)},
legend cell align={left},
legend style={font=\scriptsize, fill opacity=0.8, draw opacity=1, text opacity=1, draw=white!80!black},
tick align=inside,
tick pos=both,
x grid style={white!69.0196078431373!black},
xlabel={Inter-group distance [m]},
xmajorgrids,
xmin=10, xmax=670,
xtick style={color=black},
y grid style={white!69.0196078431373!black},
ylabel={SINR [dB]},
ymajorgrids,
ymin=17.7409435118418, ymax=50.2904301632311,
ytick style={color=black},
ylabel style={yshift=-0.15cm, font=\footnotesize\color{white!15!black}},
xlabel style={font=\footnotesize\color{white!15!black}}
]
\path [draw=color0, semithick]
(axis cs:40,19.2204656323595)
--(axis cs:40,19.9743548737724);

\path [draw=color0, semithick]
(axis cs:80,19.6684807683231)
--(axis cs:80,20.5074803466561);

\path [draw=color0, semithick]
(axis cs:160,20.4379153961694)
--(axis cs:160,21.0851884310376);

\path [draw=color0, semithick]
(axis cs:320,21.7428273042521)
--(axis cs:320,22.5120683062978);

\path [draw=color0, semithick]
(axis cs:640,23.469637270533)
--(axis cs:640,24.2187525075106);

\path [draw=color1, semithick]
(axis cs:40,31.2934587919346)
--(axis cs:40,32.1047486955705);

\path [draw=color1, semithick]
(axis cs:80,32.9476537981218)
--(axis cs:80,33.7074492867152);

\path [draw=color1, semithick]
(axis cs:160,35.0330606317834)
--(axis cs:160,35.8603716370266);

\path [draw=color1, semithick]
(axis cs:320,35.6892351036944)
--(axis cs:320,36.1157880981012);

\path [draw=color1, semithick]
(axis cs:640,36.6071620054869)
--(axis cs:640,37.1055801189867);

\path [draw=color2, semithick]
(axis cs:40,45.1267754385577)
--(axis cs:40,46.0003858965865);

\path [draw=color2, semithick]
(axis cs:80,45.0332253278237)
--(axis cs:80,45.5643788674559);

\path [draw=color2, semithick]
(axis cs:160,46.6160044161584)
--(axis cs:160,47.3378352991709);

\path [draw=color2, semithick]
(axis cs:320,48.1404906851197)
--(axis cs:320,48.7911295339373);

\path [draw=color2, semithick]
(axis cs:640,48.2384688344707)
--(axis cs:640,48.8109080427134);

\addplot [semithick, color0, dashed, mark=triangle, mark options={solid}, forget plot]
table {%
40 19.5974102530659
80 20.0879805574896
160 20.7615519136035
320 22.127447805275
640 23.8441948890218
};
\addplot [semithick, color1, solid, forget plot, mark options={solid}]
table {%
40 31.6991037437526
80 33.3275515424185
160 35.446716134405
320 35.9025116008978
640 36.8563710622368
};
\addplot [semithick, dashed, color2, mark=o, mark options={solid}, forget plot]
table {%
40 45.5635806675721
80 45.2988020976398
160 46.9769198576646
320 48.4658101095285
640 48.5246884385921
};
\end{axis}

\begin{axis}[
width=0.951\fwidth,
height=\fheight,
at={(0\fwidth,0\fheight)},
legend cell align={left},
legend style={font=\scriptsize,at={(0.9,1.05)}, anchor=south west, draw=white!80.0!black},
tick pos=left,
x grid style={white!69.01960784313725!black},
xmajorgrids,
xmin=0, xmax=840,
xtick style={color=black},
y grid style={white!69.01960784313725!black},
ymajorgrids,
ymin=0, ymax=810,
ytick style={color=black},
hide y axis,
hide x axis,
legend columns=6,
]

\addplot [solid, semithick, mark=o, color0, dashed, mark=triangle, mark options={solid}]
table [row sep=crcr] {%
-1 -1\\
-2 -2\\
};
\addlegendentry{$1\times1$}

\addplot [solid, semithick, color1,  mark options={solid}]
table [row sep=crcr] {%
-1 -1\\
-2 -2\\
};
\addlegendentry{$2\times2$}

\addplot [dashed, semithick, mark=o, color2,  mark options={solid}]
table [row sep=crcr] {%
-1 -1\\
-2 -2\\
};
\addlegendentry{$4\times4$}
\end{axis}

\end{tikzpicture}
    \caption{Average \gls{sinr}}
    \label{fig:s2SinrVsDist}
  \end{subfigure}
  \hfill%
\begin{subfigure}[t]{0.48\textwidth}
  \centering
  \setlength\fwidth{\columnwidth}
  \setlength\fheight{.6\columnwidth}
\begin{tikzpicture}

\definecolor{color0}{rgb}{0.12156862745098,0.466666666666667,0.705882352941177}
\definecolor{color1}{rgb}{1,0.498039215686275,0.0549019607843137}
\definecolor{color2}{rgb}{0.172549019607843,0.627450980392157,0.172549019607843}
\pgfplotsset{every tick label/.append style={font=\scriptsize}}

\begin{axis}[
  width=0.951\fwidth,
  height=\fheight,
  at={(0\fwidth,0\fheight)},
legend cell align={left},
legend style={font=\scriptsize, fill opacity=0.8, draw opacity=1, text opacity=1, at={(0.97,0.03)}, anchor=south east, draw=white!80!black},
tick align=inside,
tick pos=both,
x grid style={white!69.0196078431373!black},
xlabel={Inter-group distance [m]},
xmajorgrids,
xmin=10, xmax=670,
xtick style={color=black},
y grid style={white!69.0196078431373!black},
ylabel={PRR},
ymajorgrids,
ymin=0.805926575085836, ymax=1.00924159166258,
ytick style={color=black},
ylabel style={yshift=-0.15cm, font=\footnotesize\color{white!15!black}},
xlabel style={font=\footnotesize\color{white!15!black}}
]
\path [draw=color0, semithick]
(axis cs:40,0.815168166748415)
--(axis cs:40,0.835793833251585);

\path [draw=color0, semithick]
(axis cs:80,0.87953667604533)
--(axis cs:80,0.90331982395467);

\path [draw=color0, semithick]
(axis cs:160,0.952944877925394)
--(axis cs:160,0.964573822074606);

\path [draw=color0, semithick]
(axis cs:320,0.993384220478529)
--(axis cs:320,0.997566879521471);

\path [draw=color0, semithick]
(axis cs:640,0.999364408737396)
--(axis cs:640,0.999854191262604);

\path [draw=color1, semithick]
(axis cs:40,0.981200290145544)
--(axis cs:40,0.990280709854456);

\path [draw=color1, semithick]
(axis cs:80,0.999807594803461)
--(axis cs:80,0.999967105196539);

\path [draw=color1, semithick]
(axis cs:160,0.999939772277248)
--(axis cs:160,0.999999227722752);

\path [draw=color1, semithick]
(axis cs:320,1)
--(axis cs:320,1);

\path [draw=color1, semithick]
(axis cs:640,1)
--(axis cs:640,1);

\path [draw=color2, semithick]
(axis cs:40,0.999934045706888)
--(axis cs:40,0.999999154293112);

\path [draw=color2, semithick]
(axis cs:80,1)
--(axis cs:80,1);

\path [draw=color2, semithick]
(axis cs:160,1)
--(axis cs:160,1);

\path [draw=color2, semithick]
(axis cs:320,1)
--(axis cs:320,1);

\path [draw=color2, semithick]
(axis cs:640,1)
--(axis cs:640,1);

\addplot [semithick, color0, dashed, mark=triangle, forget plot, mark options={solid}]
table {%
40 0.825481
80 0.89142825
160 0.95875935
320 0.99547555
640 0.9996093
};
\addplot [semithick, color1, solid, forget plot, mark options={solid}]
table {%
40 0.9857405
80 0.99988735
160 0.9999695
320 1
640 1
};
\addplot [semithick, dashed, color2, mark=o, mark options={solid}, forget plot]
table {%
40 0.9999666
80 1
160 1
320 1
640 1
};
\end{axis}

\end{tikzpicture}
  \caption{Average \gls{prr}}
  \label{fig:s2PrrVsDist}
\end{subfigure}
\caption{Performance comparison as a function of the inter-group distance for different antenna configurations and \gls{mcs} 14. The UDP source rate is 10~Mbps.}
\label{fig:s2PrrSinrVsDist}
\end{figure*}
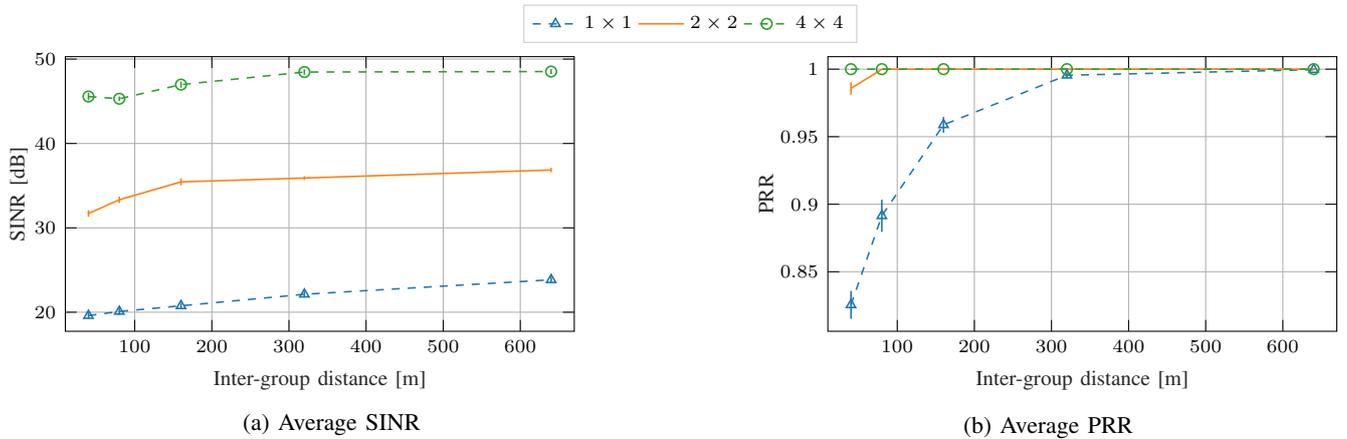

\subsection{Impact of Interference and Resource Allocation}
\label{sec:scenarioB}
In Scenario B, we considered two groups of vehicles traveling in the same direction on different lanes. Each group is composed of two vehicles, one behind the other, moving at a constant speed of $20$~m/s and keeping a safety distance of $40$~m.
Within a group, the rear vehicle acts as a server and generates data packets which are sent to the front vehicle.
We considered an ON-OFF traffic model, in which a \gls{udp} source keeps switching between the ON and the OFF states. During the ON state, the source generates packets at a constant rate for $100$~ms, while in the OFF state it stays idle for a random amount of time, which follows an exponential distribution with mean $100$~ms.
All vehicles  operate at $28$~GHz with a bandwidth of $100$~MHz, possibly interfering in case of concurrent transmissions, and are equipped with a \gls{upa} of $N\times M$ antenna elements to establish directional communications.

\begin{figure}[t!]
  \centering
  \setlength\fwidth{\columnwidth}
  \setlength\fheight{.6\columnwidth}
\begin{tikzpicture}

\definecolor{color0}{rgb}{0.12156862745098,0.466666666666667,0.705882352941177}
\definecolor{color1}{rgb}{1,0.498039215686275,0.0549019607843137}
\definecolor{color2}{rgb}{0.172549019607843,0.627450980392157,0.172549019607843}
\definecolor{color3}{rgb}{0.83921568627451,0.152941176470588,0.156862745098039}
\pgfplotsset{every tick label/.append style={font=\scriptsize}}

\begin{axis}[
  ybar,
  width=0.951\fwidth,
  height=\fheight,
  at={(0\fwidth,0\fheight)},
legend cell align={left},
legend style={font=\scriptsize, fill opacity=0.8, draw opacity=1, text opacity=1, at={(0.03,0.97)}, anchor=north west, draw=white!80!black},
tick align=inside,
tick pos=both,
x grid style={white!69.0196078431373!black},
xlabel={Offered Traffic [Mbps]},
xmajorgrids,
xtick=data,
symbolic x coords={10, 50, 100},
y grid style={white!69.0196078431373!black},
ylabel={Throughput [Mbps]},
ymajorgrids,
ytick style={color=black},
ytick={10, 50, 100},
yticklabels={10, 50, 100},
bar width=8pt,
enlarge x limits=0.3,
ylabel style={yshift=-0.15cm, font=\footnotesize\color{white!15!black}},
xlabel style={font=\footnotesize\color{white!15!black}}
]
%
%
%
%
%
%
%
%
%
%
%

\addplot [semithick, fill=red!20!white,postaction={pattern=north east lines}, fill opacity=0.8]
table {%
10 10.000000
50 38.246840
100 38.200275
};
\addlegendentry{MCS 0, Shared}
\addplot [semithick, fill=red!20!white, fill opacity=0.8]
table {%
10 10.000000
50 21.982140
100 21.869475
};
\addlegendentry{MCS 0, Dedicated}
\addplot [semithick, fill=red!80!white, fill opacity=0.8]
table {%
10 9.8230895
50 49.0274175
100 99.010760
};
\addlegendentry{MCS 28, Shared}
\addplot [semithick, fill=red!80!white,postaction={pattern=north east lines}, fill opacity=0.8]
table {%
10 10.000000
50 50.000000
100 100.000000
};
\addlegendentry{MCS 28, Dedicated}
\end{axis}

\end{tikzpicture}
  \caption{Average throughput as a function of the UDP source rate and the modulation and coding scheme, with or without the orthogonal scheduling of the resources. The antenna configuration is $2\times2$.}
  \label{fig:s2S}
\end{figure}
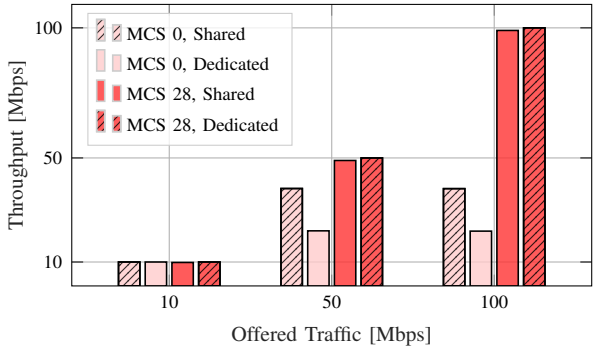

In this context, we evaluate the impact of the interference on the communication performance by considering different system configurations.
In Figure~\ref{fig:s2SinrVsDist}, we plot the average \gls{sinr} experienced as a function of the  inter-group distance, i.e., the distance between the two groups of vehicles, and of the antenna size.
It can be seen that the \gls{sinr} increases with the inter-group distance as a consequence of the weaker effect of the interference.
The trend is similar for all the antenna configurations, but the \gls{sinr} curve has a different offset depending on the number of antenna elements that are used.
Indeed, Figure~\ref{fig:s2PrrVsDist} demonstrates that  the average \gls{prr}  at the application layer increases for larger antenna arrays, which are able to focus the transmitted power on narrower beams, hence achieving a higher directivity that can possibly reduce the interference.
Specifically, only the $4\times4$ configuration is able to provide a reliable data delivery (i.e., PRR $\cong1$), regardless of the inter-group distance.
For all other antenna architectures, perfect reception is guaranteed when the inter-group distance is higher than 80 m and 600 m for the $2\times 2$ and $1\times 1$ configurations, respectively.

The presence of a centralized scheduling mechanism could prevent the occurrence of packet collisions by splitting the available resources among the groups in an orthogonal manner. However, reducing the amount of resources accessible to each terminal may limit the achievable throughput.
Moreover, the usage of a robust modulation and coding scheme, coupled with a high antenna gain, could provide protection against the interference and enable proper communication performance even without a scheduling mechanism, but at the cost of a lower capacity.
We evaluated this trade off by analyzing the average throughput achieved for  different modulation and coding schemes, either with or without the orthogonal split of the radio resources among the groups, as reported in Figure~\ref{fig:s2S}.
With \gls{mcs} 0, i.e., the most robust modulation and coding scheme, both  strategies are not always able to satisfy the offered traffic. However, in case of shared resources, the system provides a higher throughput thanks to the larger amount of available resources.
Instead, for \gls{mcs}~28 both strategies offer enough resources to
accommodate the offered traffic. We notice that the use of directional
antennas mitigates the effect of interference and guarantees high
performance even without a coordinated scheduling mechanism.
  

\section{Conclusions and Future Work}
\label{sec:concl}

In this paper we tested the  MilliCar  module that we recently released for ns-3, to evaluate the end-to-end performance of V2V networks operating at mmWaves.
We considered two simulation scenarios in which vehicles operate through directional communications to exchange data packets using a \gls{udp} application, and we investigated the impact of several system-level  parameters, including the numerology, the \gls{mcs}, the antenna array size, the RLC reordering timer, the propagation scenario, and the communication distance.
Our results demonstrated that proper beamforming design could mitigate
the effect of interference and improve the efficiency by increasing the
reuse of the available resources, thereby ensuring higher communication
performance, though at the cost of additional complexity (e.g., to align
the transmit/receive beams).
Also, we proved that, while the RLC reordering timer does not impact much on the PRR, it makes the average end-to-end delay increase significantly, especially for increasing MCSs.
Finally, we analyzed the effect of different scheduling options and we concluded that for MSC 0, i.e., modulation and coding scheme that yields the lowest datarate at the physical layer, the system provides a higher throughput when sharing the available resources, thanks to the robustness of this \gls{mcs} against errors caused by the interference.

As part of our future work, we will integrate new features to the MilliCar module based on  the latest proposals in the  3GPP NR V2X standardization process, including a more  realistic beam management mechanism and a dedicated  medium access control scheme.

\bibliographystyle{IEEEtran}
\bibliography{bibl.bib}

\end{document}